\documentclass[10pt,twocolumn,showpacs,showkeys,preprintnumbers,amssymb,aps,superscriptaddress,pre]{revtex4-2}
\usepackage{amsmath}
\usepackage{graphicx}
\usepackage{orcidlink}
\usepackage{hyperref}

\begin{document}

\preprint{APS/123-QED}

\title{Continuous and discontinuous transitions in the Ising-Heisenberg model on the extended Lieb lattice in a magnetic field}

\author{D\'avid Siv\'y\,\orcidlink{0000-0001-9486-7470}}
\affiliation{Department of Theoretical Physics and Astrophysics, Faculty of Science, 
P. J. \v{S}af\'arik University, Park Angelinum 9, 041 54 Ko\v{s}ice, Slovak Republic}
\affiliation{Institute of Physics, Bijen\'i\v{c}ka cesta 46, HR-10001 Zagreb, Croatia}
\author{Jozef Stre\v{c}ka\,\orcidlink{0000-0003-1667-6841}}
\email{Corresponding author: jozef.strecka@upjs.sk}
\affiliation{Department of Theoretical Physics and Astrophysics, Faculty of Science, 
P. J. \v{S}af\'arik University, Park Angelinum 9, 041 54 Ko\v{s}ice, Slovak Republic}

\date{\today}

\begin{abstract}
The spin-$1/2$ Ising-Heisenberg model on the extended Lieb lattice in a magnetic field is exactly mapped onto an effective spin-$1/2$ Ising model on the square lattice. The ground-state phase diagram comprises the quantum antiferromagnetic (QAF), quantum monomer–dimer (MD), classical ferrimagnetic (FRI), and classical ferromagnetic phase. The MD-FRI ground-state phase boundary extends to finite temperatures as a dome-shaped surface of discontinuous thermal transitions bounded by a line of Ising critical points. The QAF phase is enclosed by a surface of continuous thermal transitions evolving from the QAF-MD and QAF-FRI ground-state phase boundaries. Monte Carlo simulations fully confirm the existence and nature of both continuous and discontinuous thermal phase transitions obtained by exact and approximate analytical calculations.
\end{abstract}
 
\maketitle

\section{Introduction}

The study of phase transitions and critical phenomena in lattice-statistical models is fascinating topic of statistical physics, largely due to the concept of universality, which asserts that systems of very different microscopic nature may exhibit identical critical behavior in the vicinity of phase transitions \cite{fis67, wil83, sta99}. In this context, exactly solvable classical spin models exhibiting thermal phase transitions and critical points play a particularly important role as they provide rigorous insights into critical phenomena free from uncontrolled approximations \cite{dom60,dom72,bax82}. Despite their apparent simplicity, however, exactly solvable lattice-statistical Ising models that simultaneously exhibit thermal phase transitions are relatively rare and are predominantly restricted to low-dimensional systems \cite{dom60,dom72,bax82}. A paradigmatic example is the zero-field Ising model on a square lattice, whose exact solution by Onsager revealed a continuous finite-temperature phase transition accompanied by a power-law divergence of the magnetic susceptibility and a logarithmic divergence of the specific heat thereby establishing one of the most celebrated analytical benchmarks in statistical physics \cite{ons44}. 

Nevertheless, the introduction of an external magnetic field generally renders even two-dimensional Ising model non-exactly solvable despite its apparent simplicity \cite{dom60,dom72,bax82}. Another key factor governing the behavior of low-dimensional spin systems is spin frustration, which arises from competing interactions that cannot be satisfied simultaneously \cite{lie86,die04}. A prototypical example is provided by two-dimensional Ising models with competing interactions, which exhibit a variety of nontrivial phenomena originating from a geometric spin frustration \cite{lie86,die04,kal11, jin12} and still remain at the focus of ongoing research interest \cite{hwa24, roo24, cha25}. Although a full exact solution of two-dimensional Ising lattices is generally restricted to the particular case of zero magnetic field, the geometric spin frustration may relax this limitation to a certain subspace of the overall parameter space \cite{fwu85,aza88,gia88,wlu05,bar19,mut22}. 

An even greater challenge is to identify exactly solvable lattice-statistical models among low-dimensional quantum Heisenberg spin systems, which can provide invaluable benchmarks for approximate analytical methods as well as state-of-the-art numerical simulations \cite{lac11, cab12}. However, the occurrence of thermal phase transitions associated with spontaneous symmetry breaking is prohibited in low-dimensional quantum Heisenberg spin systems by the Mermin--Wagner theorem \cite{mer66}. From this perspective, exactly solved quantum Heisenberg models displaying thermal phase transitions are exceedingly rare as exact solutions are mostly restricted to one-dimensional systems \cite{mat93}. The application of an external magnetic field to two-dimensional Heisenberg models may not only enrich the magnetic behavior, but can also relax the restrictions imposed by the Mermin--Wagner theorem whose validity is limited to systems possessing a continuous spin-rotational symmetry \cite{mer66}. 

Moreover, the geometric spin frustration may significantly enrich the thermal phase transitions of two-dimensional quantum Heisenberg antiferromagnets in an external magnetic field \cite{lac11,sen04,fan24}. A paradigmatic example is provided by the fully frustrated Heisenberg bilayer, which exhibits a line of discontinuous thermal phase transitions terminating at an Ising critical point \cite{sta18}. This finding has stimulated extensive subsequent research, which proved that such behavior is a generic feature of a broader class of frustrated quantum Heisenberg antiferromagnets \cite{web22,fac22,cac23}. Although these findings may at first sight appear to be of primarily theoretical interest, an experimental realization of the spin-1/2 Heisenberg antiferromagnet on the Shastry-Sutherland lattice provided by the magnetic compound SrCu$_2$(BO$_3$)$_2$ \cite{kag99} indeed offered a unique platform for the experimental observation of a rich variety of thermal phase transitions and critical phenomena \cite{lar21,nom23,nyc25}.

One of prototypical examples of a two-dimensional frustrated quantum spin system, which displays in a magnetic field a line of finite-temperature phase transition ending at an Ising critical point, is the spin-1/2 Heisenberg antiferromagnet on a diamond-decorated square lattice \cite{cac23}.  Remarkably, the completely same type of thermal phase transitions has been identified in the analogous yet simplified spin-1/2 Ising-Heisenberg model on the diamond-decorated square lattice, where the nodal spins are replaced by classical Ising spins while the decorating spins remain fully quantum Heisenberg spins \cite{str23}.  This hybrid classical--quantum model can be mapped exactly by means of the decoration-iteration transformation (DIT) \cite{fis59, roj09, str10} onto an effective spin-1/2 Ising model on a square lattice. A particularly intriguing aspect of this model is that it remains exactly solvable in the parameter region where the discontinuous thermal phase transitions occur \cite{str23}. Specifically, a surface of discontinuous thermal phase transitions terminating at a line of Ising critical points was rigorously established in the regime of vanishing effective field, where the effective Ising model admits an exact solution though the genuine magnetic field remains finite \cite{str23}. Consequently, the exactly solved Ising-Heisenberg model provides a valuable insight into the underlying mechanism of these outstanding thermal phase transitions \cite{str23}.

This naturally raises the question of whether similar thermal phase transitions and critical behavior can also emerge in non-frustrated two-dimensional quantum spin systems in a magnetic field, in particular in the Ising-Heisenberg and Heisenberg models. Addressing this issue is important for clarifying the extent to which these phenomena are intrinsically related to a geometric spin frustration. In this context, it is desirable to identify simplified lattice-statistical models that allow for an exact analytical treatment while still capturing the essential physics of these phase transitions, since the observation of such phase transitions in exactly solvable systems may provide valuable indications of their possible occurrence in a more complex quantum spin models as demonstrated in Refs. \cite{str23, cac23}. 

The Lieb lattice represents a paradigmatic example of a bipartite lattice geometry capable of giving rise to nontrivial magnetic behavior \cite{mar55, lie62}. In particular, the extended Lieb lattice provides a convenient framework for constructing exactly solvable Ising-Heisenberg spin model with combined Ising and Heisenberg interactions \cite{str02}. Motivated by these considerations, the present work is devoted to a detailed investigation of the spin-$1/2$ Ising-Heisenberg model on the extended Lieb lattice in a magnetic field, which can be treated exactly within the framework of the DIT \cite{fis59, roj09, str10}. The resulting mapping onto an effective classical Ising model enables a transparent analysis of the phase diagram and critical behavior, while at the same time providing a useful reference point for identifying analogous phenomena in the corresponding fully quantum Heisenberg model on the same lattice.

The paper is organized as follows. In Sec.~\ref{method}, we introduce the spin-$1/2$ Ising-Heisenberg model on the extended Lieb lattice in a magnetic field and outline the main steps of the methods applied for solving it. The most important results are presented in Sec.~\ref{results} including the ground-state and finite-temperature phase diagrams along with numerical results obtained from classical Monte Carlo simulations. Finally, the main conclusions are summarized in Sec.~\ref{conclusion}.

\section{Model and method}
\label{method}

\begin{figure}[t]
\centering
\includegraphics[scale=0.07]{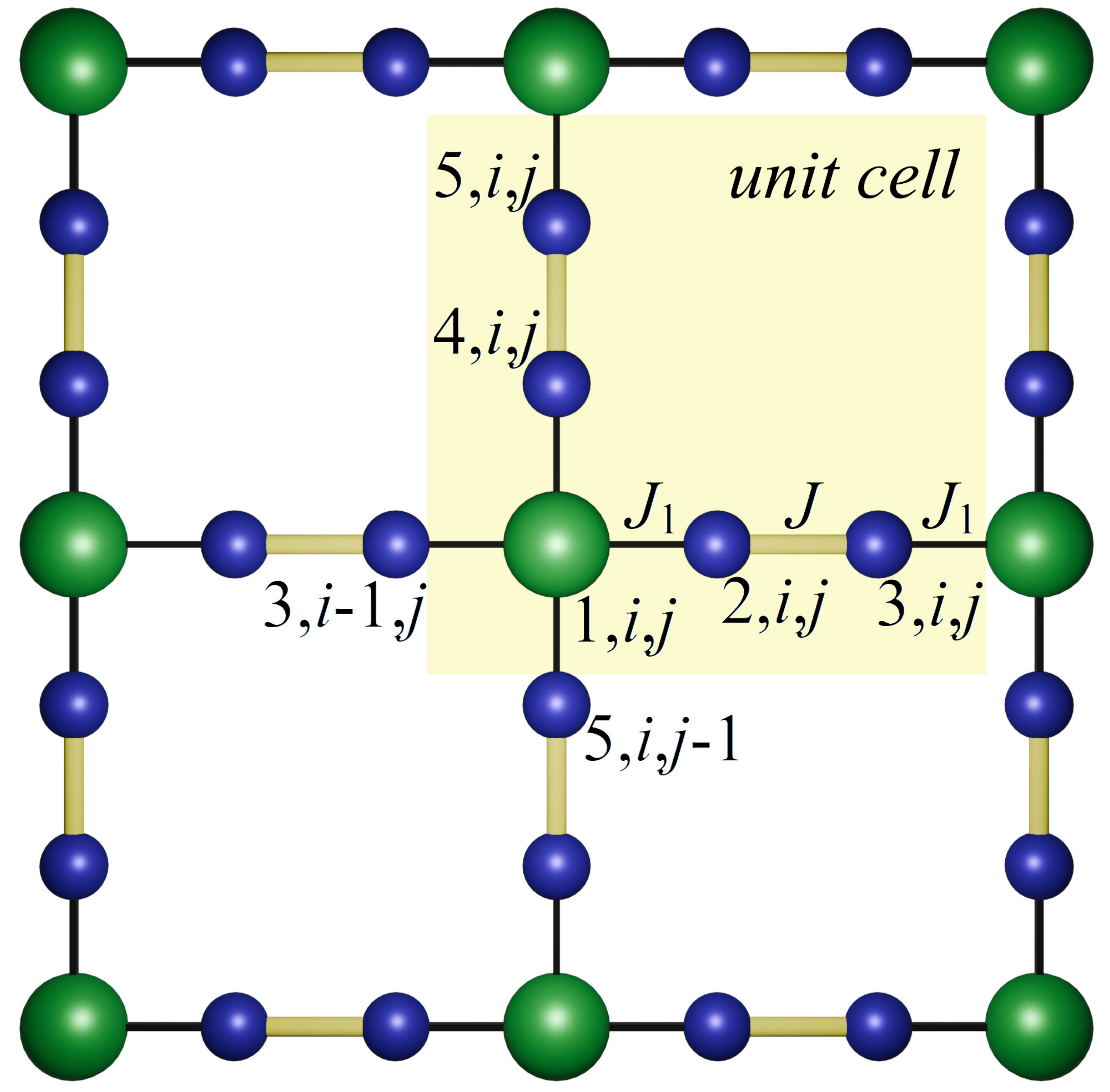}
\caption{Schematic illustration of the spin-$1/2$ Ising-Heisenberg model on the extended Lieb lattice. Green (blue) circles denote the lattice positions of the Ising (Heisenberg) spins. The five spins constituting a unit cell are highlighted by a light yellow square.}
\label{fig1}
\end{figure}

We consider the spin-$1/2$ Ising-Heisenberg model on an extended Lieb lattice schematically depicted in Fig. \ref{fig1}, which is described by the Hamiltonian
\begin{align}
\hat{\mathcal{H}} \! &= 
J_1 \sum_{i,j=1}^{L} \!
\left[ \hat{S}_{1,i,j}^{z} \left( \hat{S}_{2,i,j}^{z} + \hat{S}_{3,i-1,j}^{z} + \hat{S}_{4,i,j}^{z} + \hat{S}_{5,i,j-1}^{z} \right) \right] \nonumber \\
\! & \!+\! J  \! \sum_{i,j=1}^{L} \!\! \left(\boldsymbol{\hat{S}}_{2,i,j} \cdot  \boldsymbol{\hat{S}}_{3,i,j} \!+\! \boldsymbol{\hat{S}}_{4,i,j}  \cdot  \boldsymbol{\hat{S}}_{5,i,j} \right)  
\!-\! h \! \sum_{i,j=1}^{L} \sum_{k=1}^{5} \! \hat{S}_{k,i,j}^{z}.
\label{ham}
\end{align}
Here, the first subscript $k$ in the set of indices $(k,i,j)$ specifies the position of a spin within a unit cell, whereas the remaining two subscripts $i$ and $j$ specify the row and column indices of the corresponding unit cell. The spin-$1/2$ operators $\hat{\mathbf{S}}_{k,i,j} \equiv ( \hat{S}_{k,i,j}^{x}, \hat{S}_{k,i,j}^{y}, \hat{S}_{k,i,j}^{z})$ ($k=2,3,4,5$) are assigned to the Heisenberg spin pairs coupled via the exchange interaction $J$, which are represented in Fig.~\ref{fig1} by small blue circles. In addition, the Heisenberg spins interact with their nearest-neighbor Ising spins $\hat{S}_{1,i,j}^{z}$ represented in Fig.~\ref{fig1} by large green circles via the Ising coupling constant $J_1$. The last term in the Hamiltonian~(\ref{ham}) represents the Zeeman interaction of both the Heisenberg and Ising spins with an external magnetic field $h$.

For convenience, the total Hamiltonian (\ref{ham}) of the spin-$1/2$ Ising-Heisenberg model on the extended Lieb lattice can be decomposed into a sum of cluster Hamiltonians involving one Heisenberg spin pair and its two adjacent Ising spins 
\begin{align}
\hat{\mathcal{H}} = \sum_{i,j=1}^{L} \sum_{\delta=0}^{1}\hat{\mathcal{H}}_{i,j}^{(\delta)},
\label{ham2}
\end{align}
where the parameter $\delta \in \{0,1\}$ is introduced to unify the notation for the mutually commuting Hamiltonians $\hat{\mathcal{H}}_{i,j}^{(0)}$ and $\hat{\mathcal{H}}_{i,j}^{(1)}$ of the spin clusters residing horizontal and vertical bonds, respectively. The explicit form of the cluster Hamiltonians reads
\begin{align}
\hat{\mathcal H}_{i,j}^{(\delta)} &= 
J\,\boldsymbol{\hat S}_{2+2\delta,i,j}\cdot \boldsymbol{\hat S}_{3+2\delta,i,j} \nonumber \\
&+ J_1 \left(\hat S_{1,i,j}^{z}\hat S_{2+2\delta,i,j}^{z}+\hat S_{3+2\delta,i,j}^{z}\hat S_{1,i+1-\delta,j+\delta}^{z}
\right) \nonumber \\
&- h \! \left(\hat S_{2+2\delta,i,j}^{z}\!+\!\hat S_{3+2\delta,i,j}^{z} \right) - \frac{h}{4} \! \left(\hat S_{1,i,j}^{z}\!+\!\hat S_{1,i+1-\delta,j+\delta}^{z} \right),  \label{ham3}
\end{align}
whereby the factor $1/4$ in the last Zeeman term associated with the Ising spins prevents its overcounting as this contribution is symmetrically distributed among four different cluster Hamiltonians. The decomposition of the total Hamiltonian into the cluster Hamiltonians  (\ref{ham2}) enables a partial factorization of the partition function into the product
\begin{align}
Z &= \sum_{\{S_{1,i,j}^z\}} \prod_{i,j=1}^{L} \prod_{\delta=0}^{1}
\left[  
\operatorname{Tr}_{2+2\delta,i,j}
\operatorname{Tr}_{3+2\delta,i,j}
\exp\left(-\beta \hat{\mathcal{H}}_{i,j}^{(\delta)} \right)
\right],
\label{part}
\end{align}
where the summation extends over all possible configurations of the Ising spins and the trace is carried out over the degrees of freedom of the Heisenberg spin pairs. To  evaluate the latter trace, the cluster Hamiltonians~(\ref{ham3}) are diagonalized in the corresponding spin-dimer basis as the diagonalization procedure can be performed independently for each cluster owing to the orthogonality of the relevant Hilbert subspaces. Subsequently, the generalized DIT~\cite{fis59,roj09,str10} is employed to replace the effective Boltzmann weight associated with a given spin cluster through an equivalent expression involving less important multiplicative factor $A$, the effective interaction $J_{\rm eff}$ between the remaining Ising spins additionally experiencing the effective field $h_{\rm eff}$  
\begin{align}
&\operatorname{Tr}_{2+2\delta,i,j}\operatorname{Tr}_{3+2\delta,i,j}\exp\left(-\beta \hat{\mathcal{H}}_{i,j}^{(\delta)} \right) =
\nonumber \\
&2 \exp\left[\frac{\beta h}{4}\left(S_{1,i,j}^{z} + S_{1,i+1-\delta,j+\delta}^{z}\right)\right] \times \nonumber \\
& \Bigg\{ \! \exp \left(-\frac{\beta J}{4} \right) \cosh\left[\frac{\beta J_1}{2}\left(S_{1,i,j}^{z} + S_{1,i+1-\delta,j+\delta}^{z}\right) - \beta h\right] \nonumber \\
&+\exp\!\left(\frac{\beta J}{4}\right)\cosh \! \left[\frac{\beta}{2}\sqrt{J_1^2\left(S_{1,i,j}^{z}\!-\!S_{1,i+1-\delta,j+\delta}^{z}\right)^2\!\!+\!J^2}\right]\!\Bigg\} \nonumber \\
\! &= \!\! A \! \exp\!\Big[\beta J_{\rm eff} S_{1,i,j}^z S_{1,i+1-\delta,j+\delta}^z\!\!+\!\!\frac{\beta h_{\rm eff}}{4}\!\left(S_{1,i,j}^z\!\!+\!\!S_{1,i+1-\delta,j+\delta}^z\right)\!\!\Big]\!.
\label{mapping}
\end{align}
The mapping parameters $A$, $J_{\rm eff}$, and $h_{\rm eff}$ are determined by the self-consistency condition of DIT (\ref{mapping}) by considering all four possible combinations of the two Ising spins involved therein $S_{1,i,j}^z$ and $S_{1,i+1-\delta,j+\delta}^z$ 
\begin{align}
A &= 2 \mathrm{e}^{-\frac{\beta J}{4}} \left(V_1 V_2 V_3^2 \right)^{\frac{1}{4}}, 
\label{eqA} \\
\beta J_{\rm eff} &= \ln \left(\frac{V_1 V_2}{V_3^2} \right), 
\label{eqJeff} \\
\beta h_{\rm eff} &= \beta h - 2 \ln \left(\frac{V_1}{V_2} \right),
\label{eqheff}
\end{align}
where the following auxiliary functions $V_1-V_3$ were introduced in order to express the mapping parameters in a more compact form
\begin{align}
V_1 &= \cosh \Bigg( \frac{\beta J_1}{2} + \beta h \Bigg) 
      + \mathrm{e}^{\frac{\beta J}{2}} \cosh \left( \frac{\beta J}{2} \right), \nonumber \\
V_2 &= \cosh \Bigg( \frac{\beta J_1}{2} - \beta h \Bigg) 
      + \mathrm{e}^{\frac{\beta J}{2}} \cosh \left( \frac{\beta J}{2} \right), \nonumber \\
V_3 &= \cosh \left( \beta h \right) 
      + \mathrm{e}^{\frac{\beta J}{2}} \cosh \Bigg( \frac{\beta}{2} \sqrt{J_1^2 + J^2} \Bigg).
\label{mapping2}
\end{align}
The partition function (\ref{part}) of the spin-$1/2$ Ising-Heisenberg model on the extended Lieb lattice defined by the Hamiltonian (\ref{ham}) can thus be expressed with the help of DIT (\ref{mapping}) in terms of the partition function of an effective spin-$1/2$ Ising model on a square lattice
\begin{align}
Z(\beta,J,J_1,h)=A^{2N}Z_{\rm eff}(\beta,J_{\rm eff},h_{\rm eff}),
\label{part2}
\end{align}
which corresponds to the following effective Hamiltonian
\begin{align}
\mathcal{H}_{\rm eff} \!=\! -J_{\rm eff} \!\!\! \sum_{i,j=1}^{L} \!\!\! \left(S_{1,i,j}^z S_{1,i+1,j}^z \!+\! S_{1,i,j}^z S_{1,i,j+1}^z \right) 
\!-\! h_{\rm eff} \!\!\! \sum_{i,j=1}^{L} \!\!\! S_{1,i,j}^z\!.
\label{hamEff}
\end{align}

The exact mapping relation (\ref{part2}) established between the partition functions implies that any thermal phase transition of the original spin-$1/2$ Ising-Heisenberg model on the extended Lieb lattice belongs to the Ising universality class, since the mapping parameter $A$ given by Eq. (\ref{eqA}) is a regular function without any mathematical singularity, which means that any nonanalytic behavior must follow from singularities of the partition function of the effective Ising model. Consequently, solving the effective spin-$1/2$ Ising model on the square lattice described by Eq.(\ref{hamEff}) eventually provides the respective solution also for the spin-$1/2$ Ising-Heisenberg model on the extended Lieb lattice. Two particular cases should nevertheless be distinguished. In the former case, the vanishing effective field $h_{\rm eff}=0$ admits obtaining an exact solution for the spin-$1/2$ Ising-Heisenberg model on the extended Lieb lattice from the celebrated Onsager's solution \cite{ons44}. By contrast, the latter case with a nonzero effective field $h_{\rm eff}\neq 0$ does not admit an exact analytical solution for the effective Ising model and must therefore be treated by some powerful numerical method such as classical Monte Carlo simulations. 

First, we examine the special case of a vanishing effective field $h_{\rm eff}=0$ admitting obtaining exact results from the effective spin-1/2 Ising square lattice irrespective of whether the effective interaction is ferromagnetic ($J_{\rm eff}>0$) or antiferromagnetic ($J_{\rm eff}<0$). According to Eq.~(\ref{eqheff}), the condition of zero effective field is equivalent to the following constraint for the spin-$1/2$ Ising-Heisenberg model on the extended Lieb lattice
\begin{align}
\mathrm{e}^{\frac{\beta h}{2}} = \frac{\cosh \Bigl( \frac{\beta J_1}{2} + \beta h \Bigr) + \mathrm{e}^{\frac{\beta J}{2}} \cosh \Bigl( \frac{\beta J}{2} \Bigr)}{\cosh \Bigl( \frac{\beta J_1}{2} - \beta h \Bigr) + \mathrm{e}^{\frac{\beta J}{2}} \cosh \Bigl( \frac{\beta J}{2} \Bigr)},
\label{con}
\end{align}
which delimits the parameter region where the exact solution is available. Although the condition of zero effective field $h_{\rm eff}=0$ is trivially satisfied for arbitrary temperature when the genuine magnetic field vanishes $h=0$, the transcendental equation (\ref{con}) also admits a nontrivial solution corresponding to zero effective field $h_{\rm eff}=0$ even in the presence of a nonzero actual magnetic field $h \neq 0$. It is worth recalling that the Onsager's exact solution of the (effective) spin-$1/2$ Ising model on the square lattice \cite{ons44} generically exhibits a continuous thermal phase transition, which emerges at the (inverse) critical temperature $\beta_c J_{\rm eff}= \pm 2 \ln (1+\sqrt{2})$ with the plus (minus) sign corresponding to the ferromagnetic (antiferromagnetic) effective interaction $J_{\rm eff}>0$ ($J_{\rm eff}<0$). Both possibilities must be taken into account, since the effective interaction (\ref{eqJeff}) may change its character depending on the interplay among the coupling constants, magnetic field, and temperature. Consequently, the critical condition for the spin-$1/2$ Ising-Heisenberg model on the extended Lieb lattice constrained within the particular parameter subspace determined by the condition (\ref{con}) takes the form
\begin{align}
\! \! \! \mathrm{e}^{\frac{\beta_c h}{4}}(1\!+\!\!\sqrt{2})^{\pm 1} \!=\! \frac{\cosh \Bigl( \frac{\beta_c J_1}{2} \!+\! \beta_c h \Bigr) \!+\! \mathrm{e}^{\frac{\beta_c J}{2}} \cosh \Bigl(\frac{\beta_c J}{2} \Bigr)}{\cosh \Bigl(\!  \beta_c h \! \Bigr)\!  \! +\!  \mathrm{e}^{\frac{\beta_c J}{2}}\!  \cosh \Bigl(\!  \frac{\beta_c}{2} \sqrt{J_1^2 + J^2}  \Bigr)}.
\label{critCon}
\end{align}
The sign ambiguity $\pm$ in the above critical condition stems from the critical behavior of the effective Ising model with either ferromagnetic ($J_{\rm eff}>0$) or antiferromagnetic ($J_{\rm eff}<0$) effective interaction. Hence, the spin-$1/2$ Ising-Heisenberg model on the extended Lieb lattice undergoes a continuous phase transitions exactly at the critical temperature determined by either of the two critical conditions given by Eq. (\ref{critCon}). Furthermore, discontinuous thermal phase transitions are expected to occur in the spin-$1/2$ Ising-Heisenberg model on the extended Lieb lattice within the parameter subspace (\ref{con}) satisfying the condition of vanishing effective field $h_{\rm eff} = 0$ provided that the effective interaction is ferromagnetic $J_{\rm eff} > 0$ and temperature is set below its critical value, i.e. $\beta J_{\rm eff} > 2 \ln (1+\sqrt{2})$.  

While the effective ferromagnetic Ising square lattice with $J_{\rm eff}>0$ does not undergo any thermal phase transition in the presence of a nonzero effective field $h_{\rm eff} \neq 0$, the effective antiferromagnetic Ising square lattice with $J_{\rm eff}<0$ may still display a continuous thermal phase transition even in the presence of nonzero effective field $h_{\rm eff} \neq 0$. Although this particular parameter subspace is not exactly solvable anymore, a simple yet sufficiently accurate approximation for the critical temperature of the continuous phase transition of the antiferromagnetic Ising square lattice can be obtained  
from the critical condition proposed by M\"{u}ller-Hartmann and Zittartz \cite{mull77}
\begin{align}
\cosh \left( \frac{\beta_c h_{\rm eff}}{2} \right) = \sinh^2 \left( \frac{\beta_c J_{\rm eff}}{2} \right).
\label{critCon2}
\end{align} 
Inserting the effective interaction (\ref{eqJeff}) and effective field (\ref{eqheff}) into the approximate critical condition (\ref{critCon2}) and solving the resulting equation allows one to determine the critical points associated with continuous thermal phase transitions of the spin-$1/2$ Ising-Heisenberg model on the extended Lieb lattice in the regime of nonzero effective field $h_{\rm eff} \neq 0$. In the limiting case of a vanishing genuine magnetic field $h = 0$ necessarily implying zero effective field $h_{\rm eff} = 0$, the approximate solution for the critical points (\ref{critCon2}) reduces to the exact critical condition of the effective Ising square lattice (\ref{critCon}) 
\begin{align}
(1 + \sqrt{2})^{\pm 1} =\frac{\cosh\!\left(\frac{\beta_c J_1}{2}\right)+ \mathrm{e}^{\frac{\beta_c J}{2}} \cosh\!\left(\frac{\beta_c J}{2}\right)}{1 + \mathrm{e}^{\frac{\beta_c J}{2}} \cosh\!\left(\frac{\beta_c}{2}\sqrt{J_1^2 + J^2}\right)}.
\label{critCon3}
\end{align}

Exact mapping theorems developed by Barry and coworkers \cite{bar88, kha90, bar91, bar95} imply that any local observables involving only the Ising spins of the spin-$1/2$ Ising-Heisenberg model on the extended Lieb lattice, such as the single-site magnetization $m_{\rm I} \equiv \langle \hat{S}^z_{1,i,j} \rangle$ or the nearest-neighbor pair correlation function $\varepsilon_{\rm I} \equiv \langle \hat{S}^z_{1,i,j}\hat{S}^z_{1,i+1,j} \rangle$, can be expressed directly in terms of the corresponding quantities of the effective Ising model
\begin{align}
m_{\rm I} &\equiv \langle \hat{S}^z_{1,i,j} \rangle = \langle \hat{S}^z_{1,i,j} \rangle_{\rm eff} \equiv m_{\rm eff}, 
\nonumber \\
\varepsilon_{\rm I} &\equiv \langle \hat{S}^z_{1,i,j}\hat{S}^z_{1,i+1,j} \rangle 
= \langle \hat{S}^z_{1,i,j}\hat{S}^z_{1,i+1,j} \rangle_{\rm eff} \equiv \varepsilon_{\rm eff}.
\label{miei}
\end{align}
Following the procedure elaborated in Ref. \cite{str23}, the single-site magnetization of the Heisenberg spins can be then determined by means of the generalized Callen-Suzuki identity \cite{cal63,suz65,bal02}
\begin{align}
m_{\rm H} \!\equiv\!  \langle \hat{S}^z_{k,i,j} \rangle &  \!=\!  \frac{1}{8} \left( \frac{W_1}{V_1} \!+\!  \frac{W_2}{V_2} \!+\!  2\frac{W_3}{V_3} \right) 
\!-\!  \frac{m_{\rm I}}{2}  \left( \frac{W_1}{V_1} \!-\! \frac{W_2}{V_2} \right) \nonumber \\
& \!+\! \frac{\varepsilon_{\rm I}}{2} \left( \frac{W_1}{V_1} \!+\! \frac{W_2}{V_2} \!-\! 2\frac{W_3}{V_3} \right), \, (k \!=\!2, ..., 5).\!
\label{locHeis}
\end{align}
Here, we introduced the coefficients $W_1 = \sinh [\beta(J_1/2 + h)]$, $W_2 = \sinh[\beta(h - J_1/2)]$, and $W_3 = \sinh(\beta h)$. The total single-site magnetization is then given by $m_{\rm T} = (m_{\rm I} + 4 m_{\rm H})/5$. Accurate numerical results for the local and total magnetization of the spin-$1/2$ Ising-Heisenberg model on the extended Lieb lattice were then extracted from the corresponding results for the magnetization and pair correlation function of the effective spin-1/2 Ising model on a square lattice. To this end, we exploited classical Monte Carlo simulations of the effective spin-$1/2$ Ising model on the square lattice by employing the standard Metropolis algorithm \cite{met53} combined with a checkerboard update scheme.  Simulations were carried out for linear lattice sizes up to $L = 100$ under periodic boundary condition by performing $8 \times 10^5$ Monte Carlo steps.

\section{Results and discussion}
\label{results}

In this section, we present the most interesting results for the spin-$1/2$ Ising-Heisenberg model on the extended Lieb lattice in an external magnetic field. We begin with a discussion of the ground-state phase diagram and subsequently provide a detailed analysis of the finite-temperature properties including the global phase diagram, phase transitions, magnetization curves, and magnetic susceptibility.

\subsection*{A. Ground-state properties}

\begin{figure*}[t]
\centering
\begin{minipage}[t]{0.36\textwidth}
    \centering
    \includegraphics[width=\linewidth]{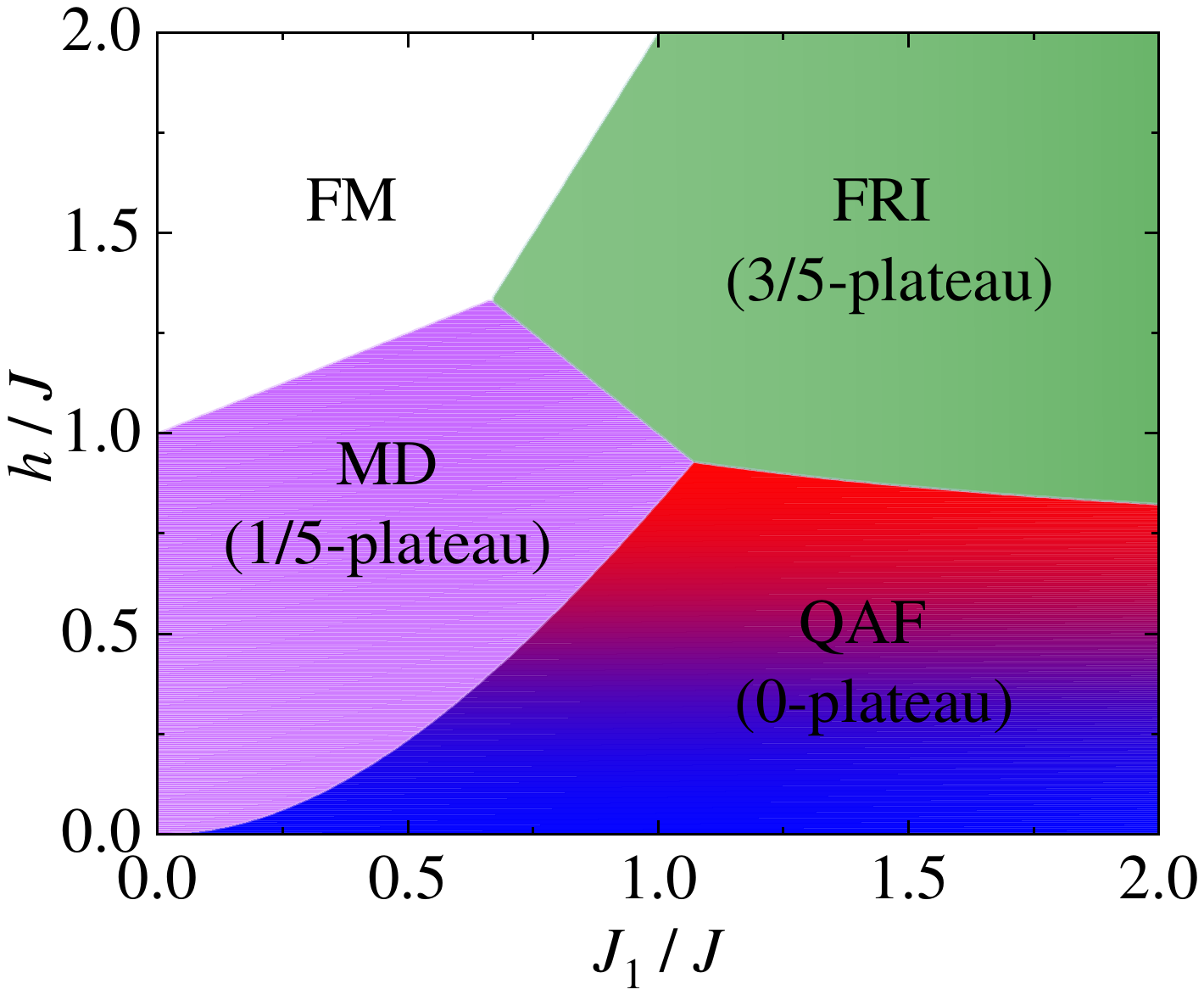}
\end{minipage}
\hfill
\begin{minipage}[t]{0.63\textwidth}
    \centering
    \includegraphics[width=\linewidth, trim=10 -200 0 0, clip]{Fig2b.pdf}
\end{minipage}
\vspace*{-0.3cm}
\caption{Ground-state phase diagram of the spin-$1/2$ Ising-Heisenberg model on the extended Lieb lattice in the $J_{1}/J$--$h/J$ plane together with schematic illustrations of the quantum antiferromagnetic (QAF), the quantum monomer-dimer (MD), and the classical ferrimagnetic (FRI) phases. Arrows indicate the spin orientation. FM denotes the fully saturated ferromagnetic phase. A violet oval represents a singlet-dimer state, whereas a double-colored oval indicates a singlet-dimer-like state.}
\label{fig2}
\end{figure*}

By comparing energies of all possible eigenstates of the spin-$1/2$ Ising-Heisenberg model on the extended Lieb lattice one finds four distinct ground states given by the following eigenvectors
\begin{align}
|{\rm QAF} \rangle  \!&=\! \prod_{i,j=1}^{L/2} |\!\uparrow_{1,2i-1,2j-1}\rangle|\!\uparrow_{1,2i,2j}\rangle|\!\downarrow_{1,2i-1,2j}\rangle|\!\downarrow_{1,2i,2j-1}\rangle \nonumber \\
&\qquad \quad\otimes \left(a_{+}|\!\uparrow_{2,i,j}\downarrow_{3,i,j}\rangle- a_{-}|\!\downarrow_{2,i,j}\uparrow_{3,i,j}\rangle\right) \nonumber \\
&\qquad \quad \otimes \left(a_{+}|\!\uparrow_{4,i,j}\downarrow_{5,i,j}\rangle- a_{-}|\!\downarrow_{4,i,j}\uparrow_{5,i,j}\rangle\right),
\label{GroundState1} \\
|{\rm MD} \rangle \!&=\! \prod_{i,j=1}^L |\!\uparrow_{1,i,j}\rangle \otimes 
\frac{1}{\sqrt{2}}(|\!\uparrow_{2,i,j}\downarrow_{3,i,j}\rangle - |\!\downarrow_{2,i,j}\uparrow_{3,i,j}\rangle) \nonumber \\
&\qquad \quad \otimes \frac{1}{\sqrt{2}}(|\!\uparrow_{4,i,j}\downarrow_{5,i,j}\rangle - |\!\downarrow_{4,i,j}\uparrow_{5,i,j}\rangle), 
\label{GroundState2} \\
|{\rm FRI} \rangle \!&=\! \prod_{i,j=1}^L  |\downarrow_{1,i,j}\rangle \otimes |\uparrow_{2,i,j}\uparrow_{3,i,j}\rangle \otimes |\uparrow_{4,i,j}\uparrow_{5,i,j}\rangle,
\label{GroundState3} \\
|{\rm FM} \rangle \!&=\! \prod_{i,j=1}^L|\!\uparrow_{1,i,j}\rangle \otimes|\!\uparrow_{2,i,j}\uparrow_{3,i,j}\rangle \otimes|\!\uparrow_{4,i,j}\uparrow_{5,i,j}\rangle,
\label{GroundState4}
\end{align}
where $a_{\pm} = \frac{1}{\sqrt{2}} (1 \pm J_1/\sqrt{J_1^2 + J^2})$ and the abbreviations QAF, MD, FRI, and FM denote the quantum antiferromagnetic, quantum monomer-dimer, classical ferrimagnetic, and classical ferromagnetic phase, respectively. Exact expressions for the ground-state phase boundaries corresponding to discontinuous phase transitions between these ground states are obtained by comparing the respective energies:
\begin{align}
&h_{\rm QAF-FRI} = \frac{2}{3} \left(\sqrt{J^2+J_1^2} + J - J_1\right), \nonumber \\ 
&h_{\rm QAF-MD} = 2 \left(\sqrt{ J^2+J_1^2 } - J\right), \nonumber \\
&h_{\rm MD-FRI} = 2J-J_1, \,
h_{\rm MD-FM} = J + \frac{J_1}{2}, \, 
h_{\rm FRI-FM} = 2J_1.
\label{boundaries}
\end{align}
Fig. \ref{fig2} displays the overall ground-state phase diagram in the $J_{1}/J$–$h/J$ plane, which comprises four distinct ground states (\ref{GroundState1})–(\ref{GroundState4}) separated by discontinuous zero-temperature phase transitions determined by the conditions (\ref{boundaries}) together with schematic illustrations of the QAF, MD, and FRI phases. The most remarkable spin arrangement occurs in the QAF phase (\ref{GroundState1}), where the Heisenberg spin pairs form dimer-singlet-like state (double-colored ovals) further mediating an effective antiferromagnetic interaction between the Ising spins displaying perfect N\'eel long-range order. Hence, the QAF phase is characterized by an intriguing coexistence of quantum and classical antiferromagnetic long-range order on the sublattice of the Heisenberg and Ising spins, respectively. By contrast, the MD phase (\ref{GroundState2}) consist of perfect dimer-singlet states of the Heisenberg spin pairs (violet ovals) accompanied by fully polarized Ising spins. This spin arrangement originates from the almost complete effective decoupling of the Ising spins induced by the singlet states of the Heisenberg dimers, which allows the Ising spins to fully align with the external magnetic field. The MD phase should consequently give rise to an intermediate 1/5-plateau in the zero-temperature magnetization curve due to the polarized nature of the Ising spins and zero contribution of the Heisenberg dimers. Additionally, the classical ferrimagnetic spin arrangement can be detected within the FRI phase (\ref{GroundState3}) in which the Heisenberg spin pairs are fully polarized along the magnetic-field direction, whereas the Ising spins are aligned in opposite to the field direction. As a result, the FRI phase should give rise to an intermediate 3/5-plateau in the zero-temperature magnetization curve due to an incomplete cancellation of the contributions of the Heisenberg and Ising spins. Finally, the systems enters at sufficiently high magnetic fields exceeding the saturation value the trivial fully polarized ferromagnetic phase FM with all Ising and Heisenberg spins aligned parallel to the external magnetic field. It should be emphasized that two triple points, at which three distinct ground states coexist, emerge at $J_1/J = 2/3$, $h/J = 4/3$ and $J_1/J = \frac{2}{3}(\sqrt{13}-2) \approx 1.07$, $h/J = \frac{2}{3}(5-\sqrt{13}) \approx 0.93$. These two triple points delimit the ground-state phase boundary between the MD and FRI phases (see Fig. \ref{fig2}).

\begin{figure}[t]
\centering
\includegraphics[width=0.45\textwidth]{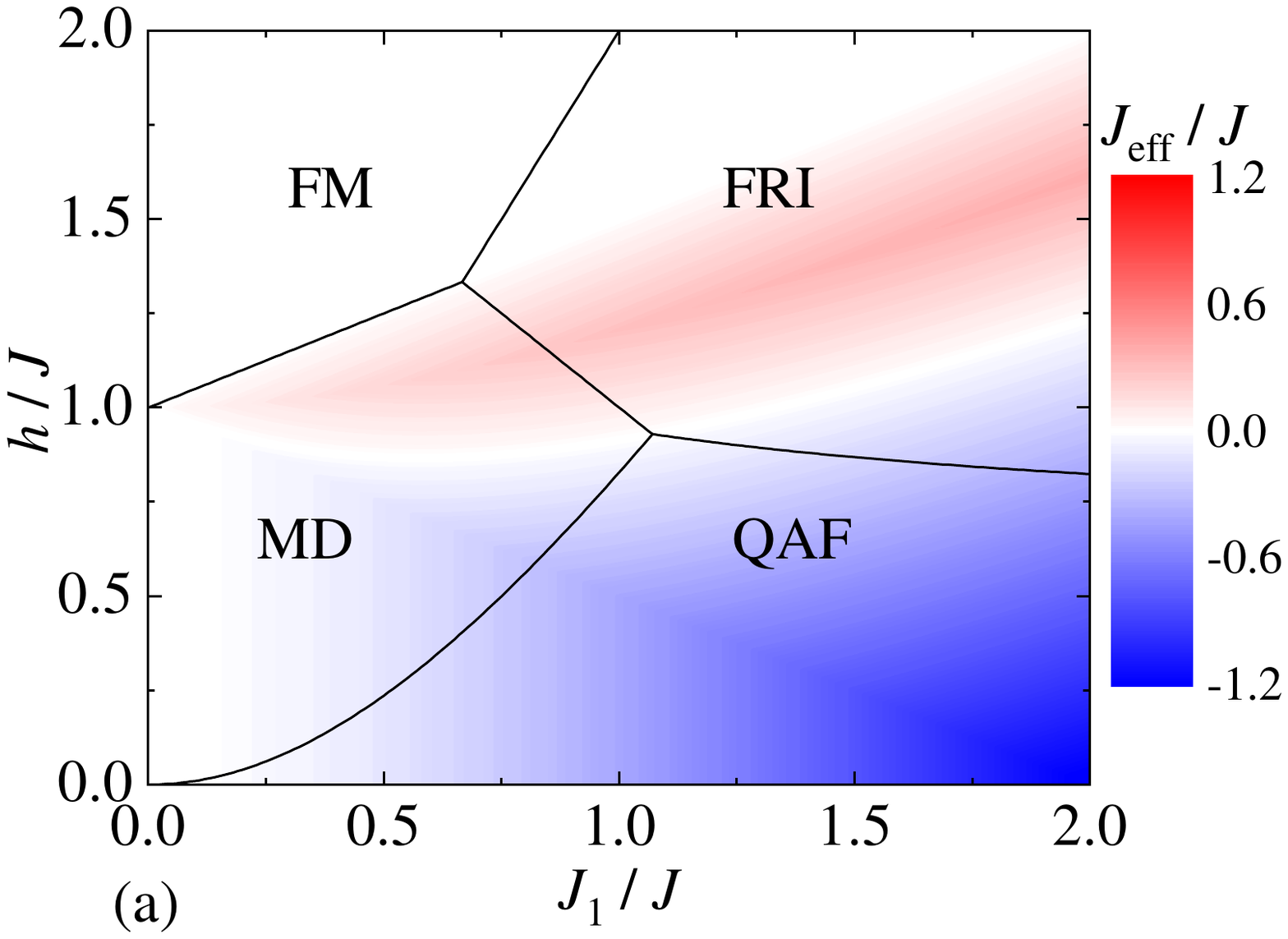}
\hspace{-0.2cm}
\includegraphics[width=0.45\textwidth]{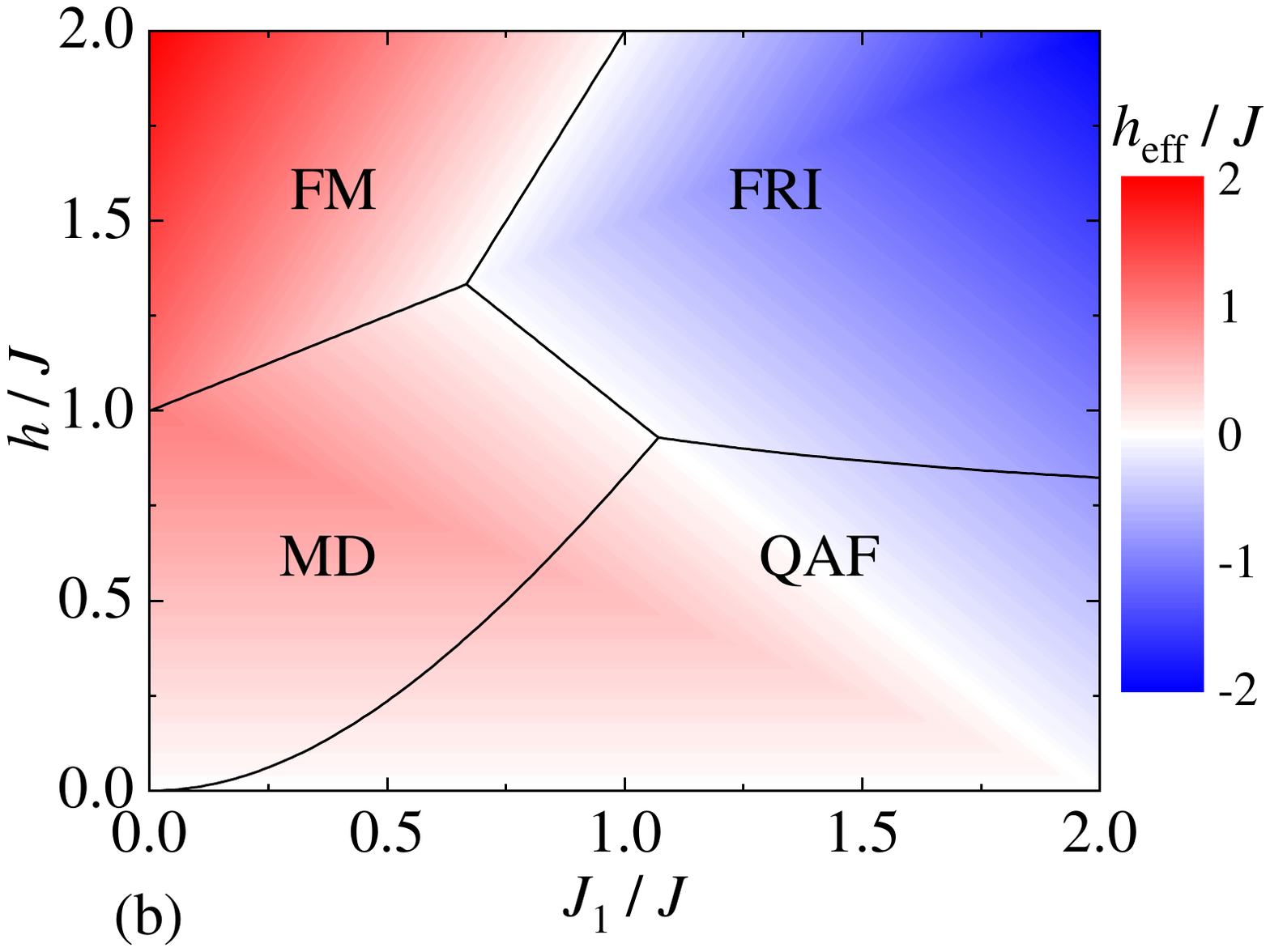}
\hspace{-0.2cm}
\vspace*{-0.5cm}
\caption{Density plots of zero-temperature asymptotic values of the effective interaction $J_{\rm eff}$ (a) and the effective field $h_{\rm eff}$ (b) of the spin-$1/2$ Ising-Heisenberg model on the extended Lieb lattice in the $J_{1}/J$--$h/J$ plane.}
\label{fig3}
\end{figure}

To provide deeper insight into the exact mapping onto the effective Ising model (\ref{hamEff}), it is quite instructive to examine the zero-temperature asymptotic values of the effective interaction $J_{\rm eff}$ (\ref{eqJeff}) and the effective field $h_{\rm eff}$ (\ref{eqheff}), which are plotted in Fig. \ref{fig3} as density plots in the $J_{1}/J$–$h/J$ plane together with the relevant ground-state phase boundaries (\ref{boundaries}) of the spin-$1/2$ Ising-Heisenberg model on the extended Lieb lattice. A necessary condition for the occurrence of finite-temperature (thermal) phase transitions in the effective Ising model (\ref{hamEff}), and consequently also in the original Ising-Heisenberg model (\ref{ham}), is a nonzero effective interaction $J_{\rm eff} \neq 0$. Therefore, one may anticipate the absence of thermal phase transitions above both ground-state boundaries associated with the FM phase [see Fig. \ref{fig3}(a)]. On the other hand, Fig. \ref{fig3} reveals that the ground-state phase boundary between the MD and FRI phases lies entirely within the parameter region characterized by a nonzero ferromagnetic effective interaction $J_{\rm eff}>0$ and a vanishing effective field $h_{\rm eff}=0$, which should contrarily support the existence of thermal phase transitions emerging above this ground-state boundary. Another important feature apparent from Fig. \ref{fig3}(a) is that the effective interaction becomes antiferromagnetic  $J_{\rm eff} < 0$ within the parameter region predominantly occupied by the QAF phase. This observation suggests the possible emergence of thermal phase transitions in this particular parameter space, since the effective Ising model with the antiferromagnetic interactions may display finite-temperature phase transitions both in zero as well as nonzero effective field [see Fig. \ref{fig3}(b)]. While the effective field $h_{\rm eff}$ changes sign within the parameter region corresponding to the QAF phase, the MD phase lies entirely in the parameter region with $h_{\rm eff}>0$, which implies a complete alignment of the Ising spins along the magnetic-field direction. By contrast, the FRI phase is entirely located in the parameter region with $h_{\rm eff}<0$ indicating that the Ising spins are fully aligned in opposite to the external magnetic field.

\subsection*{B. Thermal phase transitions}

\begin{figure}[t]
\includegraphics[width=0.6\textwidth]{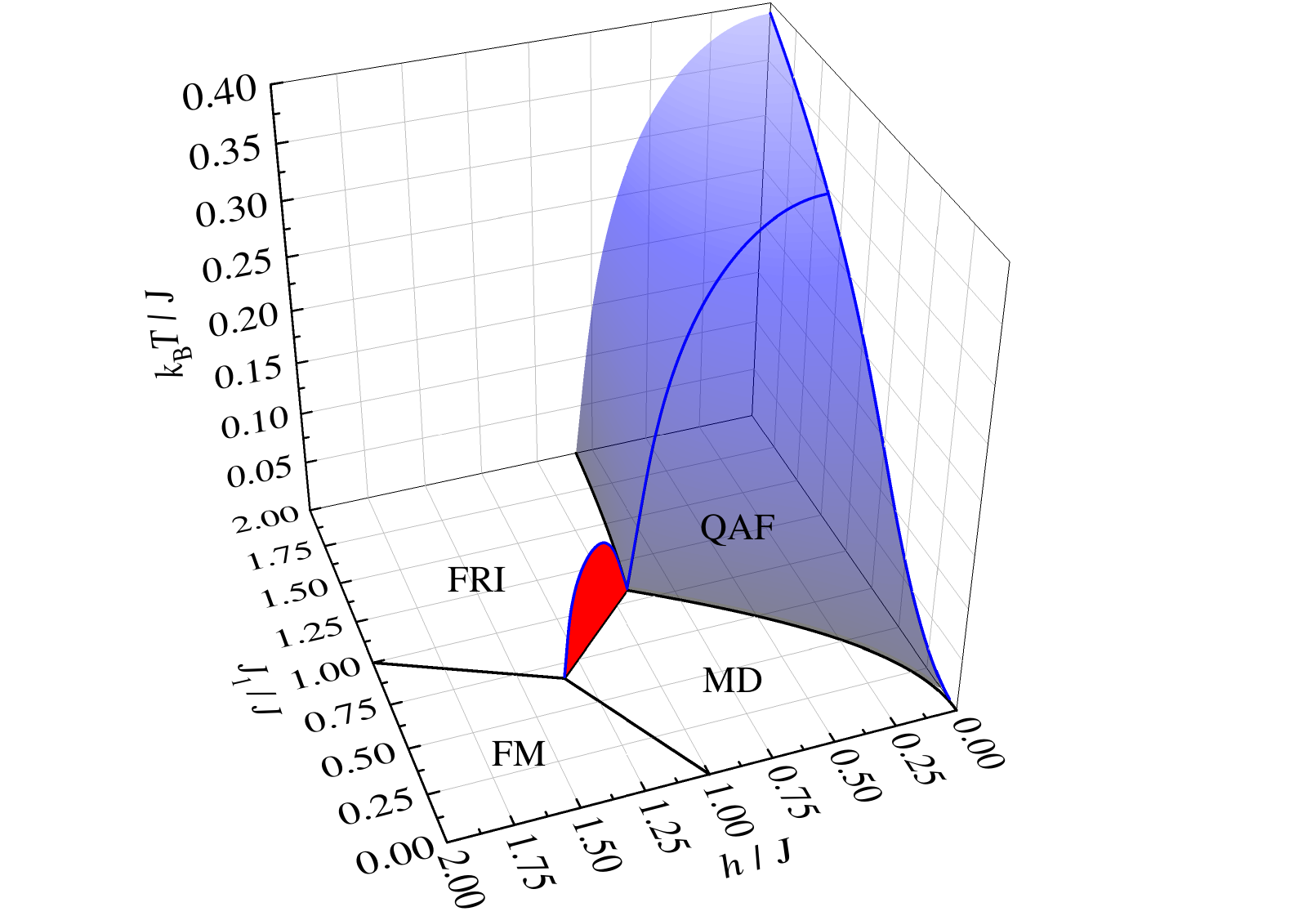}
\vspace*{-0.3cm}
\caption{Global phase diagram of the spin-$1/2$ Ising-Heisenberg extended Lieb lattice in the $J_1/J$-$h/J$-$k_{\rm B}T/J$ parameter space. The blue lines correspond to continuous phase transitions occurring at zero effective field $h_{\rm eff}=0$. The red surface represents discontinuous thermal phase transitions extending from the ground-state boundary between the MD and FRI phases and terminating at the line of continuous phase transitions. The blue surface corresponds to continuous thermal phase transitions extending from the QAF-MD and QAF-FRI ground-state phase boundaries, whereby the two surfaces continuously merge into one another at zero effective field $h_{\rm eff}=0$ as indicated by the blue solid line.}
\label{fig4}
\end{figure}

To provide a comprehensive understanding of the continuous and discontinuous thermal phase transitions of the spin-$1/2$ Ising-Heisenberg model on the extended Lieb lattice, Fig. \ref{fig4} displays its global phase diagram in a three-dimensional parameter space $J_1/J$–$h/J$–$k_{\rm B}T/J$, which extends the ground-state phase diagram discussed above. A red dome-shaped surface originating from the ground-state boundary between the MD and FRI phases corresponds to discontinuous thermal transitions between these two phases. This dome of discontinuous phase transitions was rigorously determined from the condition (\ref{con}), which ensures zero effective field $h_{\rm eff}=0$ for the effective ferromagnetic Ising square lattice with a sufficiently strong effective interaction satisfying $\beta J_{\rm eff} > 2 \ln (1 + \sqrt{2})$, i.e. a sufficiently low temperature. The dome-shaped wall of discontinuous phase transitions is bounded by a line of continuous thermal phase transitions determined from the exact critical condition (\ref{critCon}), along which the effective interaction reaches the critical value $\beta_c J_{\rm eff} = 2 \ln (1 + \sqrt{2})$ while the effective field still remains zero $h_{\rm eff}=0$ (see the blue line bounding the red surface). The global phase diagram further contains a blue critical surface associated with continuous thermal phase transitions, which relate to the thermally-assisted breakdown of the antiferromagnetic long-range order inherent to the QAF phase.  Apart from the two lines of continuous phase transitions depicted as blue solid lines, which were exactly determined from the critical condition (\ref{critCon}) for the effective antiferromagnetic interaction $J_{\rm eff}<0$ and zero effective field $h_{\rm eff}=0$, the remaining part of the critical surface was determined from the approximate critical condition (\ref{critCon2}) for the effective antiferromagnetic Ising square lattice in a nonzero effective field. The overall critical surface of continuous phase transitions actually consists of two connected critical surfaces separated by a critical line corresponding to zero effective field emerging at non-zero genuine magnetic field, which divides the critical surface associated with continuous phase transitions between the QAF and MD phases from the one associated with continuous phase transitions between the QAF and FRI phases.

\begin{figure*}[!t]
\centering
\includegraphics[width=0.45\textwidth]{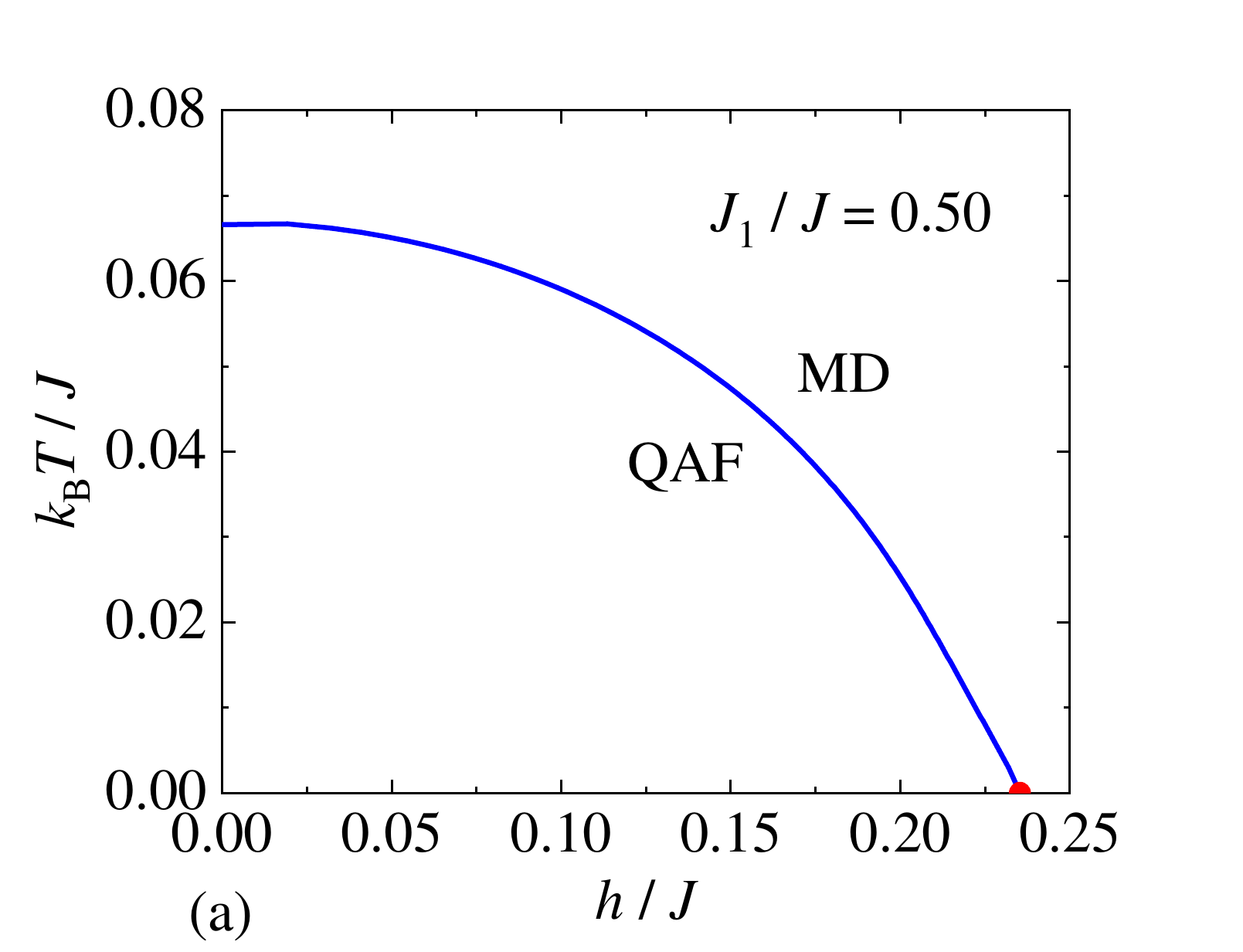}
\hspace{-0.2cm}
\includegraphics[width=0.45\textwidth]{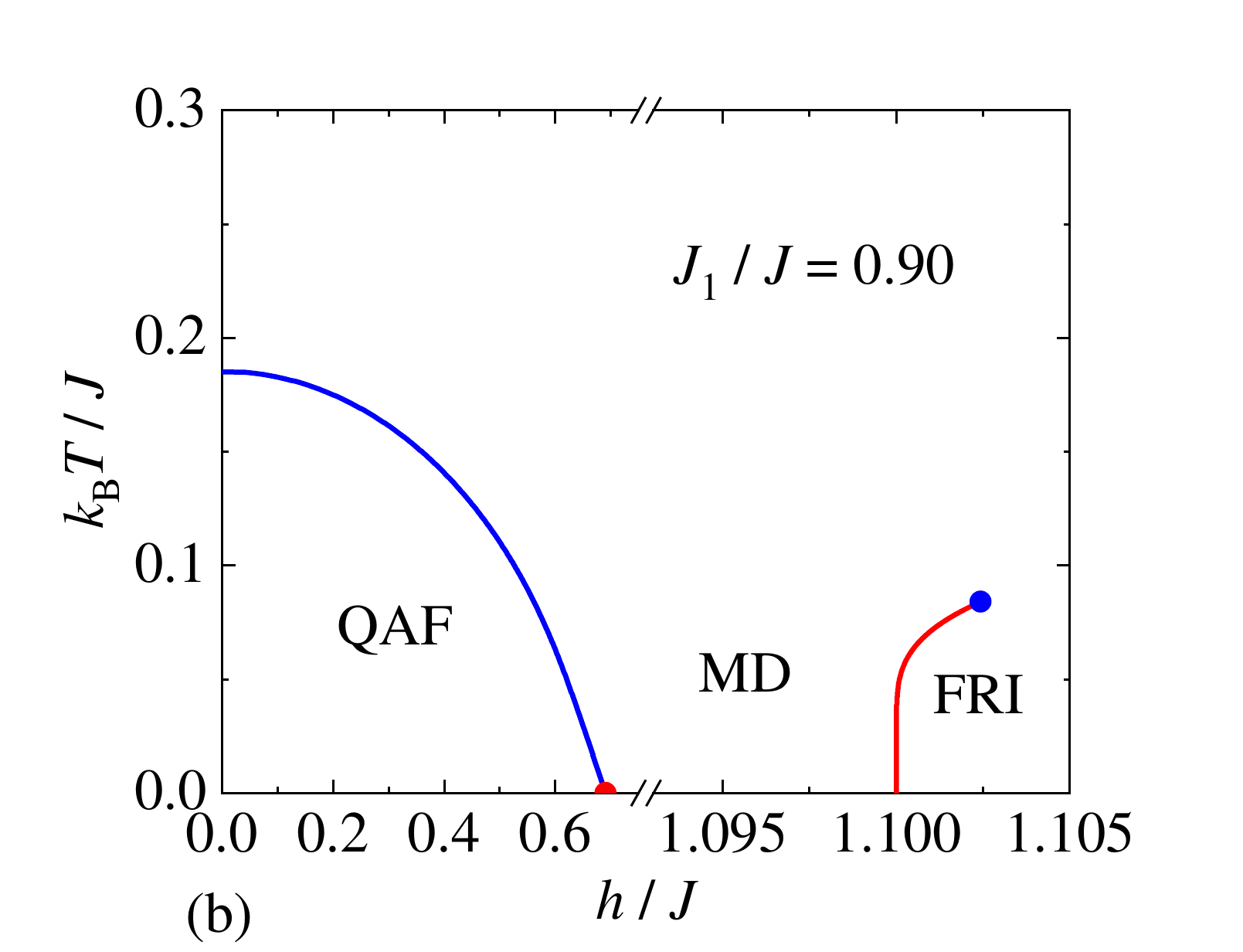}
\includegraphics[width=0.45\textwidth]{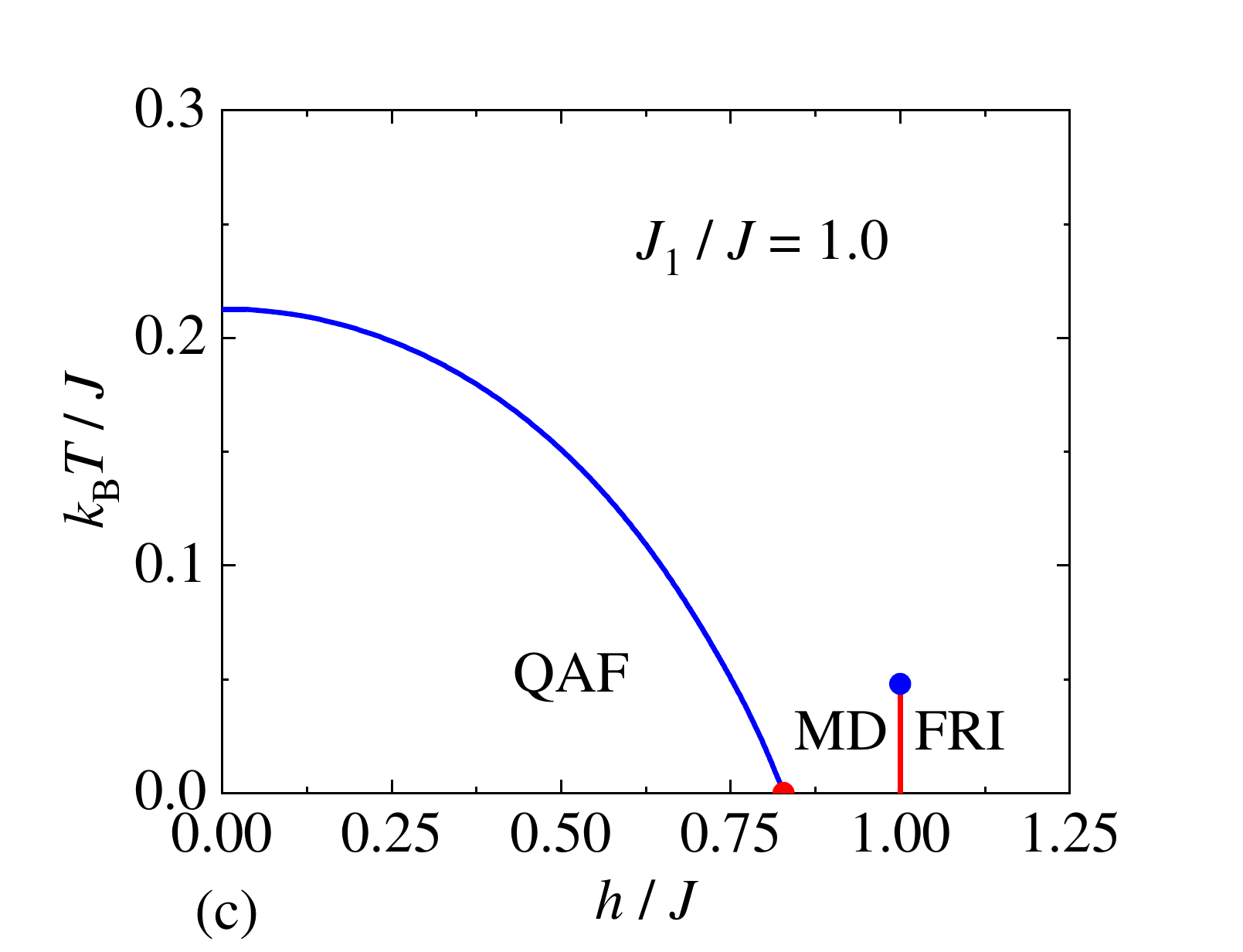}
\hspace{-0.2cm}
\includegraphics[width=0.45\textwidth]{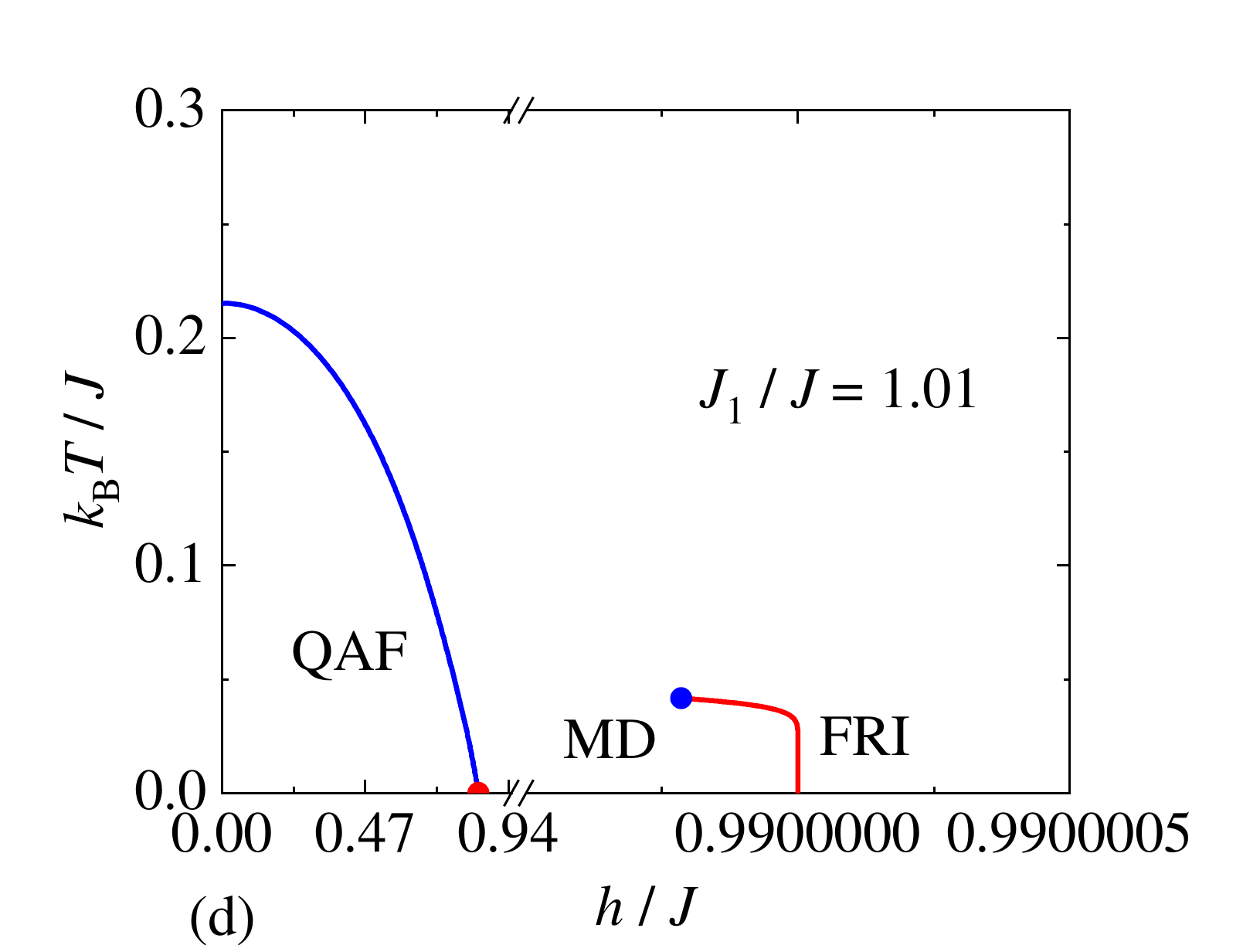}
\includegraphics[width=0.45\textwidth]{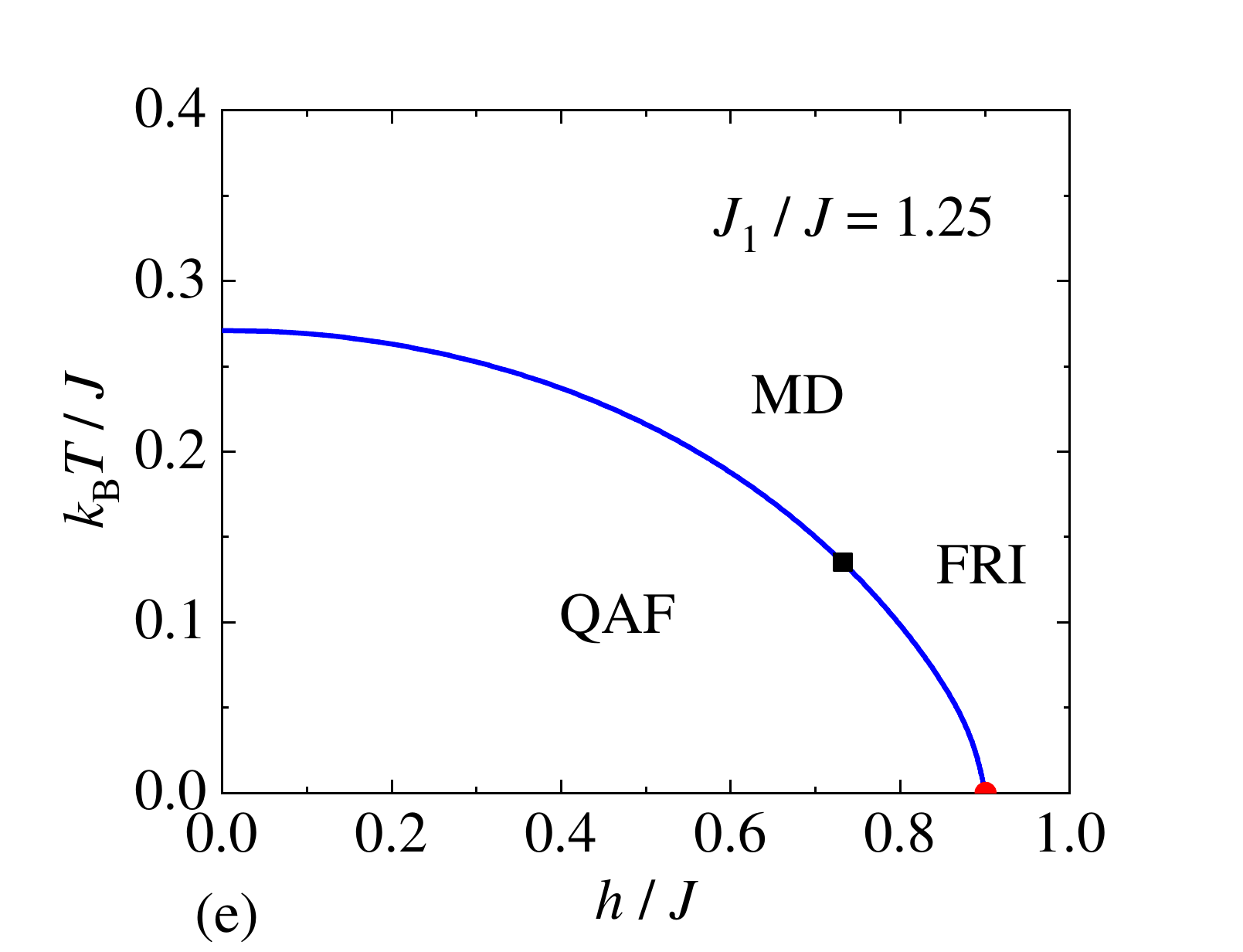}
\hspace{-0.2cm}
\includegraphics[width=0.45\textwidth]{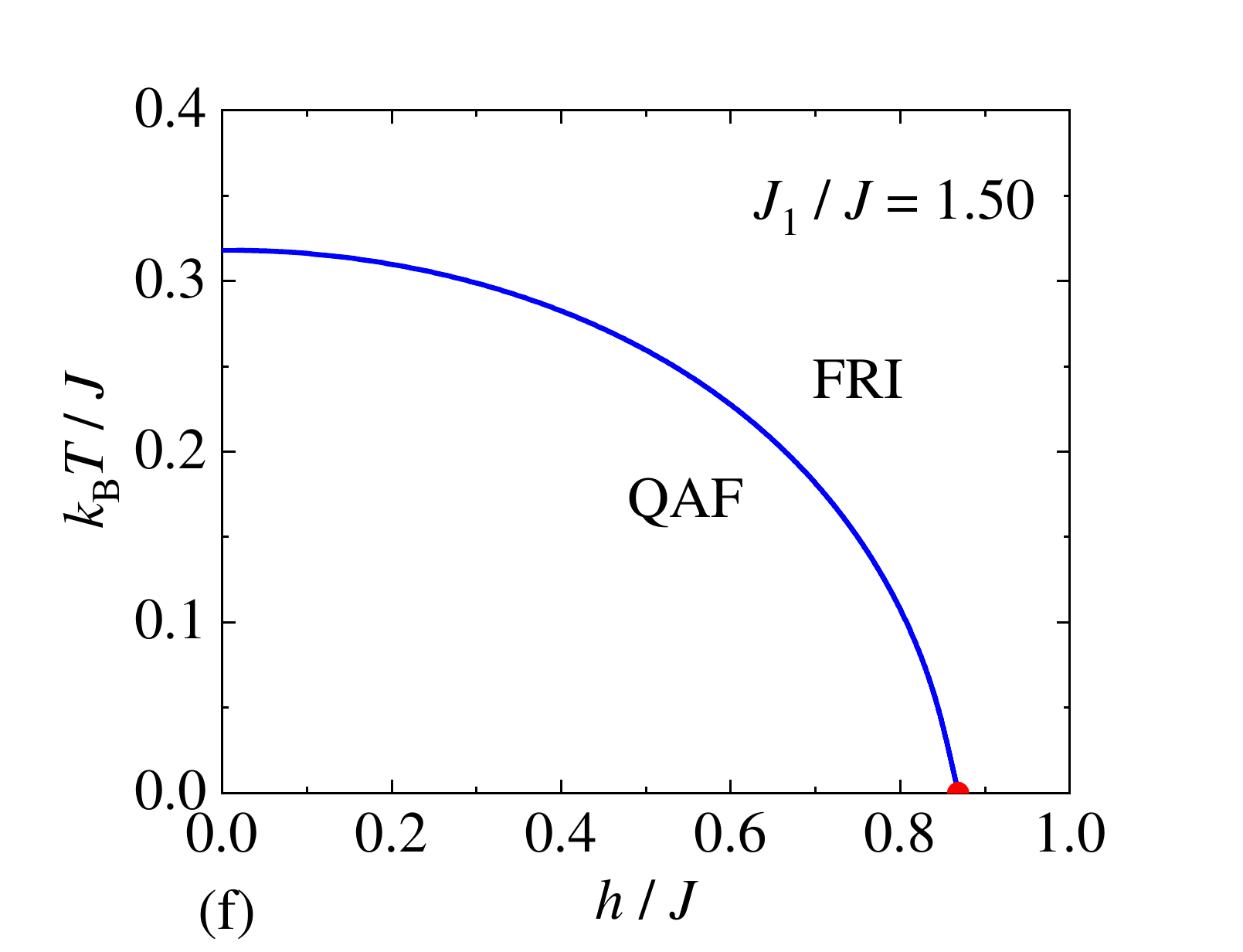}
\vspace*{-0.3cm}
\caption{Finite-temperature phase diagrams of the spin-$1/2$ Ising-Heisenberg model on the extended Lieb lattice in the $h/J$–$k_{\rm B}T/J$ plane for a few selected values of the interaction ratio: (a) $J_1/J = 0.5$, (b) $J_1/J = 0.9$, (c) $J_1/J = 1.0$, (d) $J_1/J = 1.01$, (e) $J_1/J = 1.25$, and (f) $J_1/J = 1.50$. Blue curves denote continuous phase transitions between the QAF phase and either MD or FRI phase,whereas red curves represent discontinuous transitions between the MD and FRI phases terminating at the Ising critical point shown as a blue circle. The lines of continuous phase transitions emerge from the zero-temperature discontinuous transition depicted as a red circle. The black square indicates the condition of vanishing effective field.}
\label{fig5}
\end{figure*}

For a more detailed analysis of the thermal phase transitions, Fig. \ref{fig5} shows finite-temperature phase diagrams of the spin-$1/2$ Ising-Heisenberg model on the extended Lieb lattice in the $h/J$–$k_{\rm B}T/J$ plane for six representative values of the interaction ratio $J_1/J$. Blue and red curves denote the lines of continuous and discontinuous thermal phase transitions, respectively, with the latter terminating at the Ising critical point indicated by a blue circle. Zero-temperature discontinuous phase transitions, from which the the critical lines emerge, are marked by red circles. Fig. \ref{fig5}(a) demonstrates the finite-temperature phase diagram for sufficiently small values of the interaction ratio (e.g. $J_1/J = 0.5$), which is composed only from the critical line associated with continuous phase transitions between the QAF and MD phases. For moderate values of the interaction ratio $\frac{2}{3} < J_1/J < \frac{2}{3}(\sqrt{13}-2)$, one additionally detects in the phase diagram the line of discontinuous phase transitions between the MD and FRI phases terminating at the Ising critical point [see Figs. \ref{fig5}(b)–(d)]. At the same time, the QAF phase persists up to higher temperatures and magnetic fields with increasing the interaction ratio $J_1/J$. Although the dome-like surface related to discontinuous thermal transitions between the MD and FRI phases presented in Fig. \ref{fig4} seems to be perfectly vertical, the enlarged scale used in Figs.~\ref{fig5}(b)--(d) reveals a slight but distinct inclination. Specifically, Fig. \ref{fig5}(b) illustrates that the line of discontinuous phase transitions bends towards higher fields at relatively smaller values of the interaction ratio (e.g., $J_1/J = 0.9$). By contrast, Fig. \ref{fig5}(c) indicates that the line of discontinuous phase transitions at $J_1/J = 1$ becomes almost perfectly vertical, whereas the line bends towards lower fields upon further increase of the interaction ratio as demonstrated in Fig. \ref{fig5}(d) for the interaction ratio $J_1/J = 1.01$. As the interaction ratio exceeds the interval of intermediate values, the line of discontinuous transitions between the MD and FRI phases eventually vanishes and the field range of the QAF phase gradually shrinks (see Fig. \ref{fig5}(e) for $J_1/J = 1.25$).  Moreover, Fig.~\ref{fig5}(e) demonstrates that the line of continuous phase transitions consists of two distinct segments: the low-field, high-temperature segment corresponds to the phase transitions between the QAF and MD phases, whereas the high-field, low-temperature segment corresponds to the phase transitions between the QAF and FRI phases. These two segments can be distinguished by the sign of the effective field, and therefore the condition of vanishing effective field $h_{\rm eff}=0$ indicated by the black square representing a natural criterion for distinguishing them. Last but not least, the finite-temperature phase diagram for even higher values of the interaction ratio such as $J_1/J = 1.5$ contains only the line of continuous transitions between the QAF and FRI phases [see Fig. \ref{fig5}(f)].

\subsection*{C. Monte Carlo simulations}

To provide an independent confirmation of the finite-temperature phase diagrams presented above, we extracted accurate numerical results for the spin-$1/2$ Ising-Heisenberg model on the extended Lieb lattice described by the Hamiltonian (\ref{ham}) from classical Monte Carlo simulations of the corresponding effective spin-1/2 Ising model on the square lattice given by the Hamiltonian (\ref{hamEff}). For this purpose, we specifically considered three representative finite-temperature phase diagrams displayed in Figs. \ref{fig5}(b), \ref{fig5}(e), and \ref{fig5}(f) for the interaction ratio $J_1/J = 0.9$, $1.25$, and $1.5$, respectively.

\begin{figure*}[t]
\centering
\includegraphics[width=0.35\textwidth]{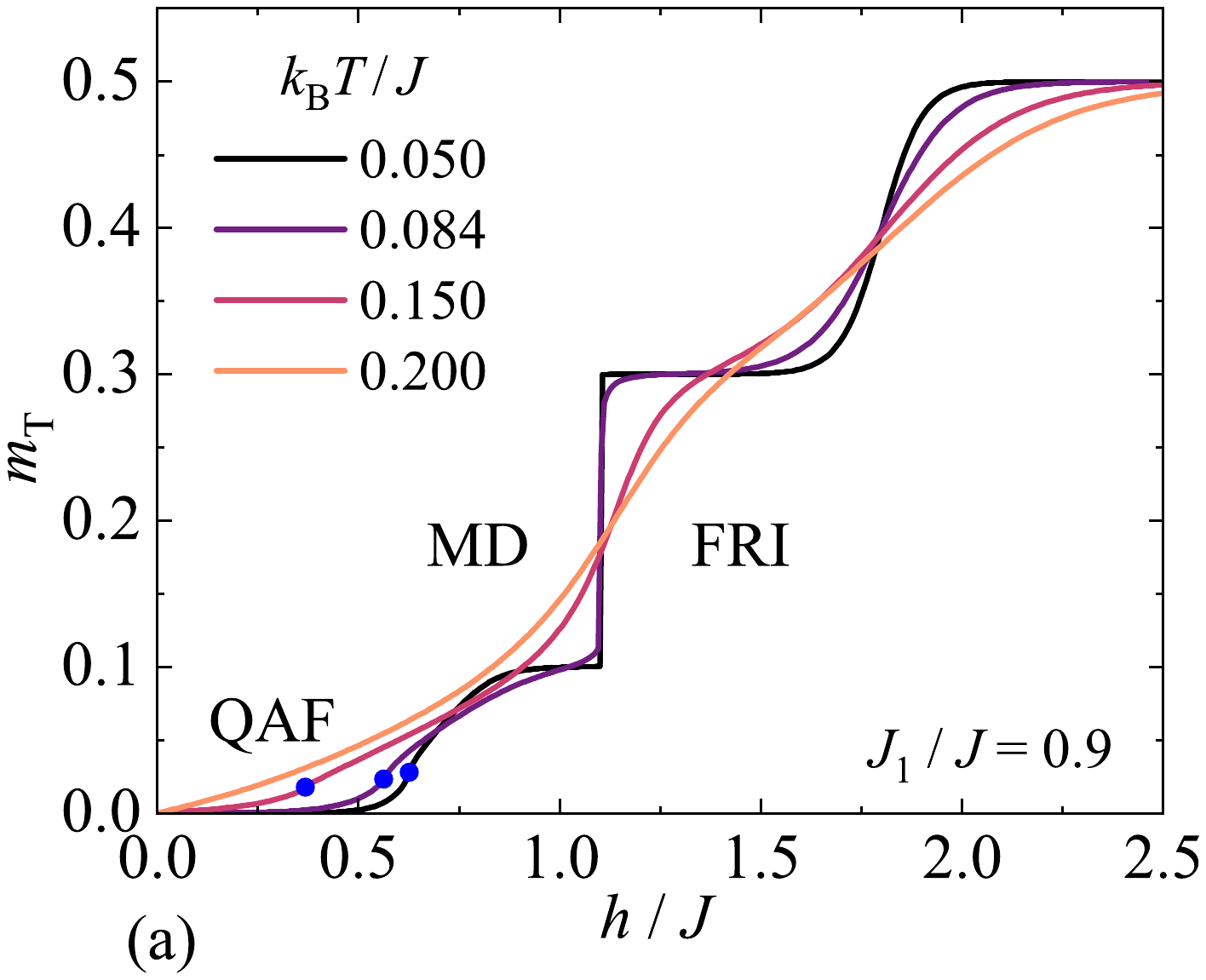}
\hspace{-0.7cm}
\includegraphics[width=0.35\textwidth]{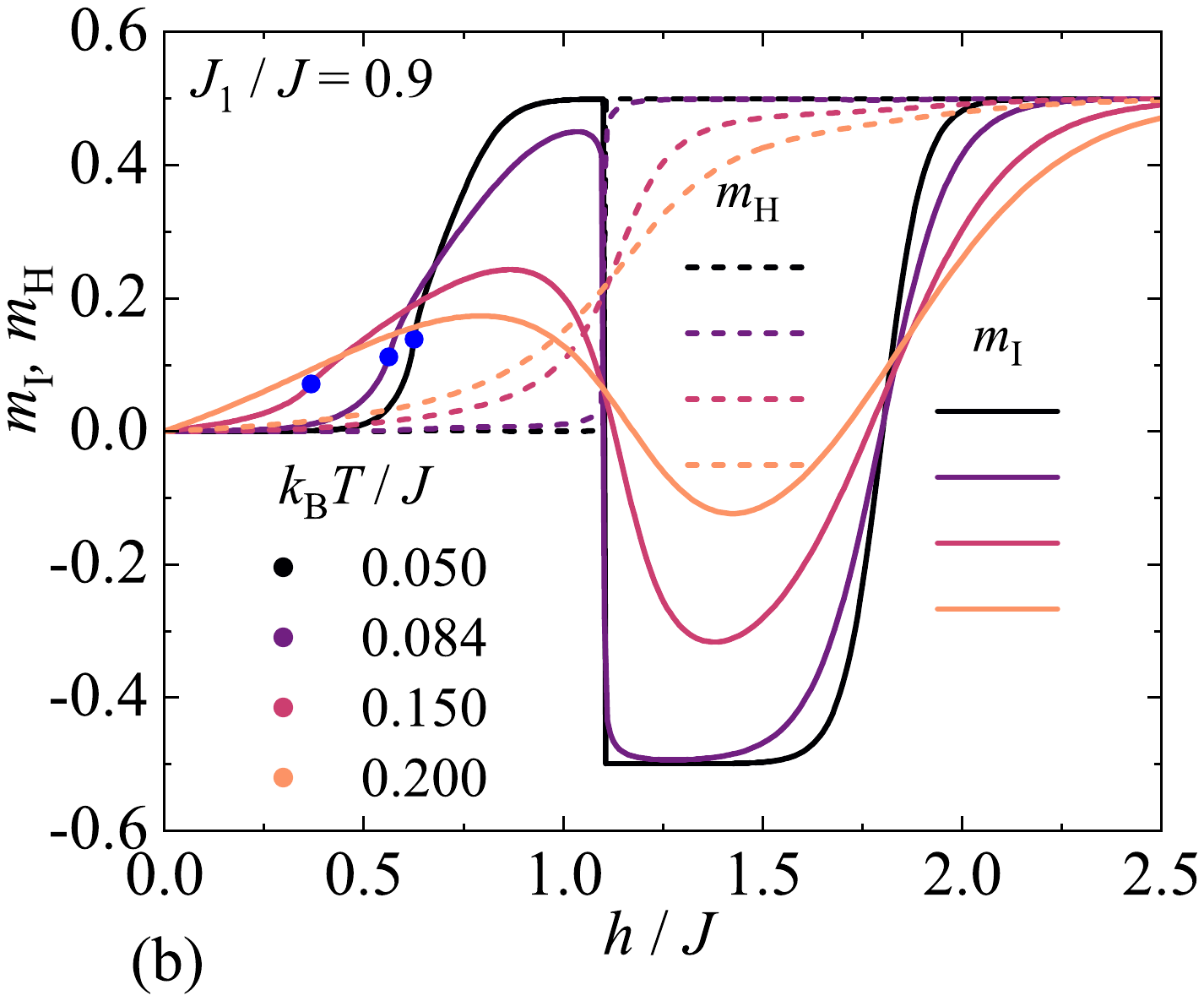}
\hspace{-0.7cm}
\includegraphics[width=0.35\textwidth]{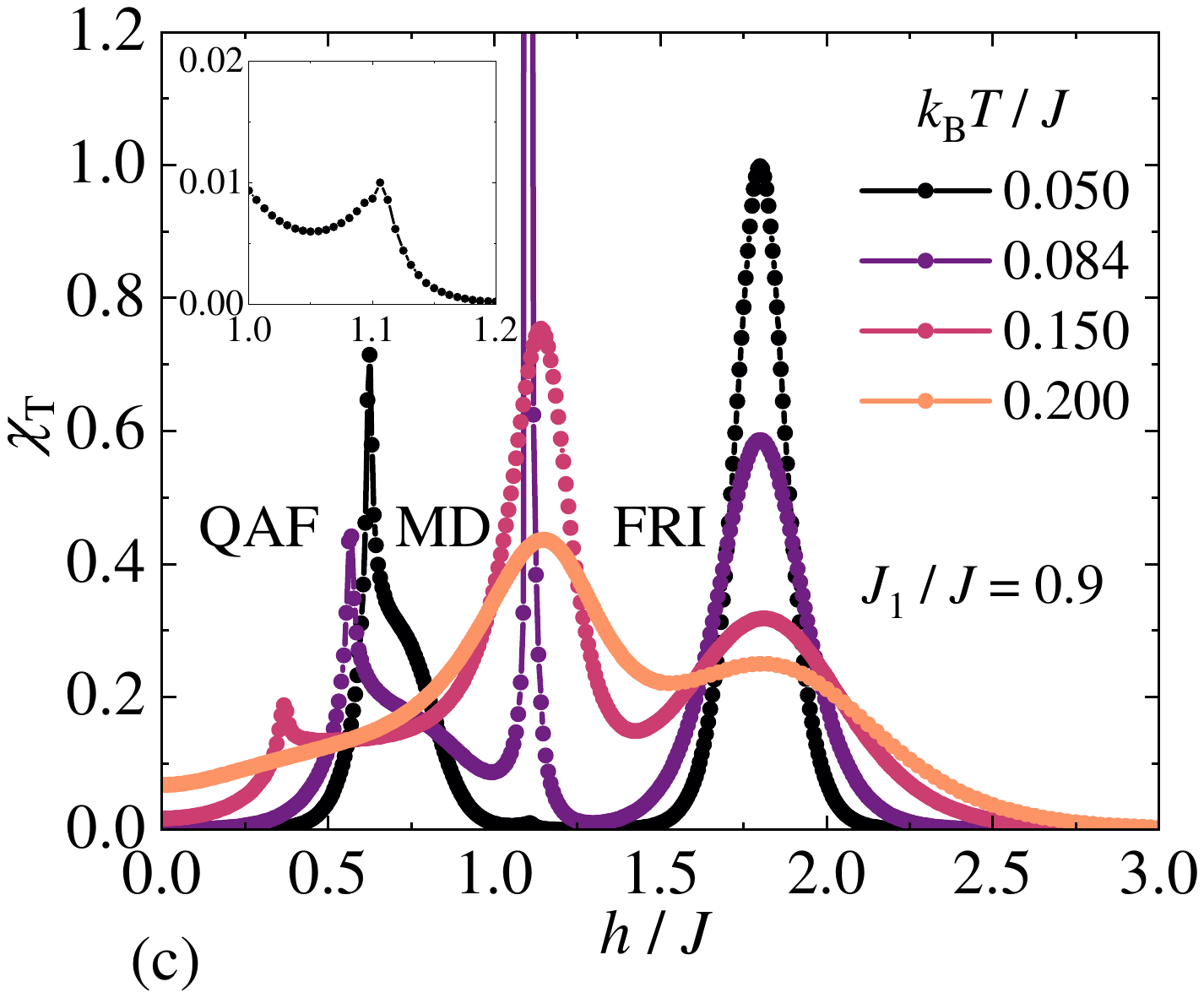}
\vspace*{-0.3cm}
\caption{Magnetic-field dependencies of the total magnetization (a), local magnetizations (b), and total magnetic susceptibility (c) of the spin-$1/2$ Ising-Heisenberg model on the extended Lieb lattice for the interaction ratio $J_1/J = 0.9$ at four different temperatures. Blue circles highlight the Ising-type critical points between the QAF and MD phases, which are omitted from the local magnetization of the Heisenberg spins for better clarity.}
\label{fig6}
\end{figure*}

For the first verification, we selected the finite-temperature phase diagram corresponding to the interaction ratio $J_1/J = 0.9$, because it exhibits a rich variety of both continuous and discontinuous thermal phase transitions as depicted in Fig. \ref{fig5}(b). Fig. \ref{fig6}(a) displays the total single-site magnetization of the spin-$1/2$ Ising-Heisenberg model on the extended Lieb lattice for this interaction ratio at four different temperatures, where blue circles indicate the locations of the Ising critical points associated with continuous phase transitions. The magnetization curves apparently exhibit three intermediate plateaus at 0, $1/5$, and $3/5$ of the saturation magnetization corresponding to the QAF, MD, and FRI phases, respectively. The magnetization curve at the lowest considered temperature $k_{\rm B}T/J = 0.05$ exhibits a continuous transition between the QAF and MD phases at the critical field $h/J \approx 0.7$ subsequently followed by a discontinuous thermal transition between the MD and FRI phases accompanied by a finite magnetization jump emerging at the second transition field $h/J \approx 1.1$. Finally, this low-temperature magnetization curve displays a steep but continuous crossover at the higher field $h/J \approx 1.8$, which signals a zero-temperature discontinuous transition between the FRI and FM phases that does not survive as a true phase transition at  finite temperatures. With increasing temperature, the discontinuity in the magnetization curves observed near the second transition field $h/J \approx 1.1$ gradually diminishes and eventually disappears around the critical temperature $k_{\rm B}T_c/J \approx 0.084$, at which the discontinuous transition between the MD and FRI phases changes into a continuous transition associated with the existence of Ising-type critical point. Indeed, the isothermal magnetization curves for higher temperatures $k_{\rm B}T/J = 0.15$ and $0.2$ no longer exhibit any thermal phase transitions between the MD and FRI phases though the former magnetization curve at $k_{\rm B}T/J = 0.15$ still retains the continuous phase transition between the QAF and MD phases, which is evidently more robust against the thermal effects. 

The nature of the QAF, MD, and QFI phases ascribed to the observed intermediate magnetization plateaus can be unambiguously identified from the corresponding field dependencies of the local magnetization displayed in Fig. \ref{fig6}(b), where solid and dashed lines represent the local magnetizations of the Ising and Heisenberg spins, respectively. The local magnetization of the Heisenberg spins remains nearly zero at the lowest temperature $k_{\rm B}T/J = 0.05$ up to the magnetic fields $h/J \approx 1.1$, which is consistent with the dimer-singlet-like and perfect dimer-singlet character of the Heisenberg spin pairs realized within the QAF and MD phases, respectively. By contrast, Fig. \ref{fig6}(b) demonstrates that the local magnetization of the Ising spins exhibits a sudden increase at the lowest temperature in the vicinity of the Ising-type critical point $h/J \approx 0.7$, above which it gradually approaches the saturated value in agreement with the existence of the MD phase. Around the second transition field $h/J \approx 1.1$, the local magnetization of the Ising spins undergoes at sufficiently low temperatures $k_{\rm B}T/J \lesssim 0.084$ an abrupt change from positive to negative values reflecting the discontinuous transition from the MD phase to the FRI phase. This transition is simultaneously accompanied by an abrupt rise in the local magnetization of the Heisenberg spins from nearly zero to an almost fully saturated value. At higher magnetic fields, both local magnetizations become smooth continuous functions for any finite temperature. 

The numerical results derived from the classical Monte Carlo simulations are thus in a full agreement with the anticipated magnetic behavior of the spin-$1/2$ Ising-Heisenberg model on the extended Lieb lattice inferred from the corresponding finite-temperature phase diagram. To further support these findings, Fig. \ref{fig6}(c) presents the corresponding field variations of the magnetic susceptibility, which exhibits several pronounced finite or divergent peaks across the entire field range at discontinuous and continuous thermal phase transitions. Starting from zero magnetic field, Fig. \ref{fig6}(c) displays three marked peaks associated with the continuous transition between the QAF and MD phases for the three lowest temperatures considered $k_{\rm B}T/J = 0.05$, $0.084$, and $0.15$.  It should be noted that the height of these peaks increases with increasing system size and becomes divergent in the thermodynamic limit. Contrary to this, the inset of Fig. \ref{fig6}(c) displays on an enhanced scale a small finite cusp in the magnetic susceptibility emerging at the lowest temperature $k_{\rm B}T/J = 0.05$ for the second transition field $h/J \approx 1.1$, which corresponds to the discontinuous transition between the MD and FRI phases. This peak remains finite irrespective of the system size although it gradually increases with temperature and tends to diverge as the temperature approaches the critical temperature $k_{\rm B}T_c/J \approx 0.084$, at which the discontinuous transition between the MD and FRI phases changes into the continuous one. At temperatures exceeding the critical value, this peak evolves into a broad rounded finite maximum confirming the absence of the thermal phase transition. Similarly, a finite round maximum of the magnetic susceptibility appears at any nonzero temperature near the saturation field $h/J \approx 1.8$, which reflects a finite-temperature crossover behavior between the FRI and FM phases as a residual signature of the zero-temperature field-driven phase transition between them.

\begin{figure*}[t]
\centering
\includegraphics[width=0.35\textwidth]{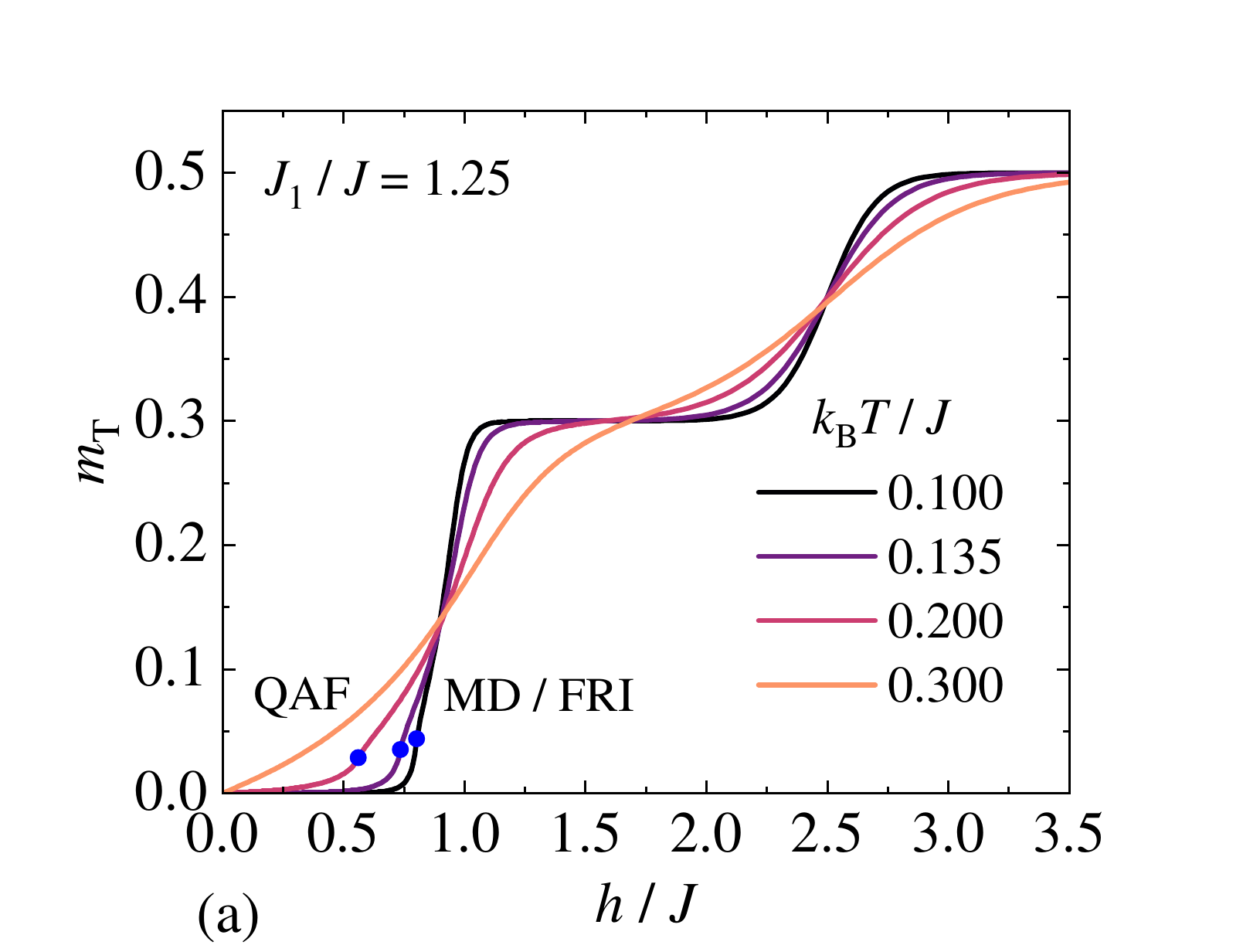}
\hspace{-0.7cm}
\includegraphics[width=0.35\textwidth]{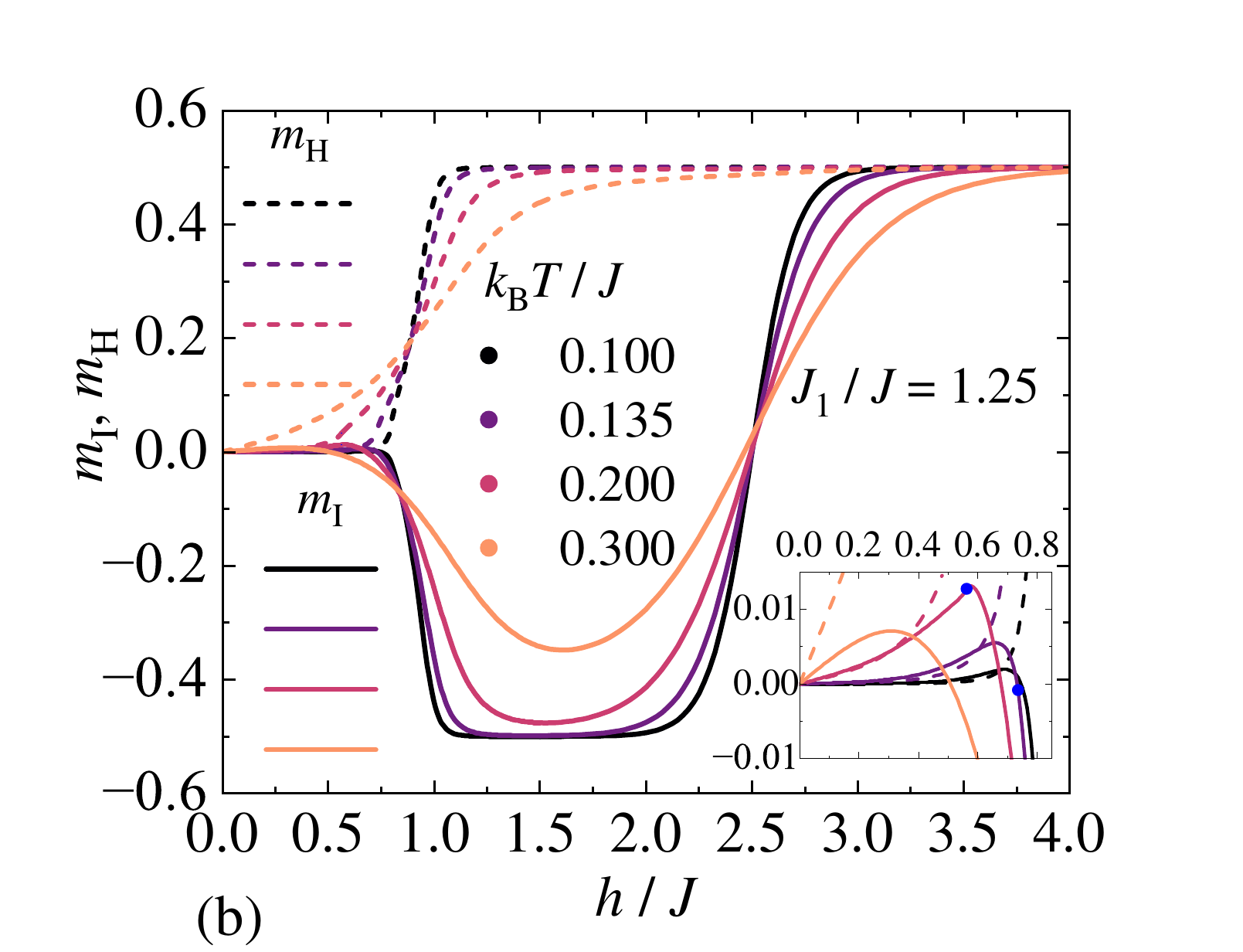}
\hspace{-0.7cm}
\includegraphics[width=0.35\textwidth]{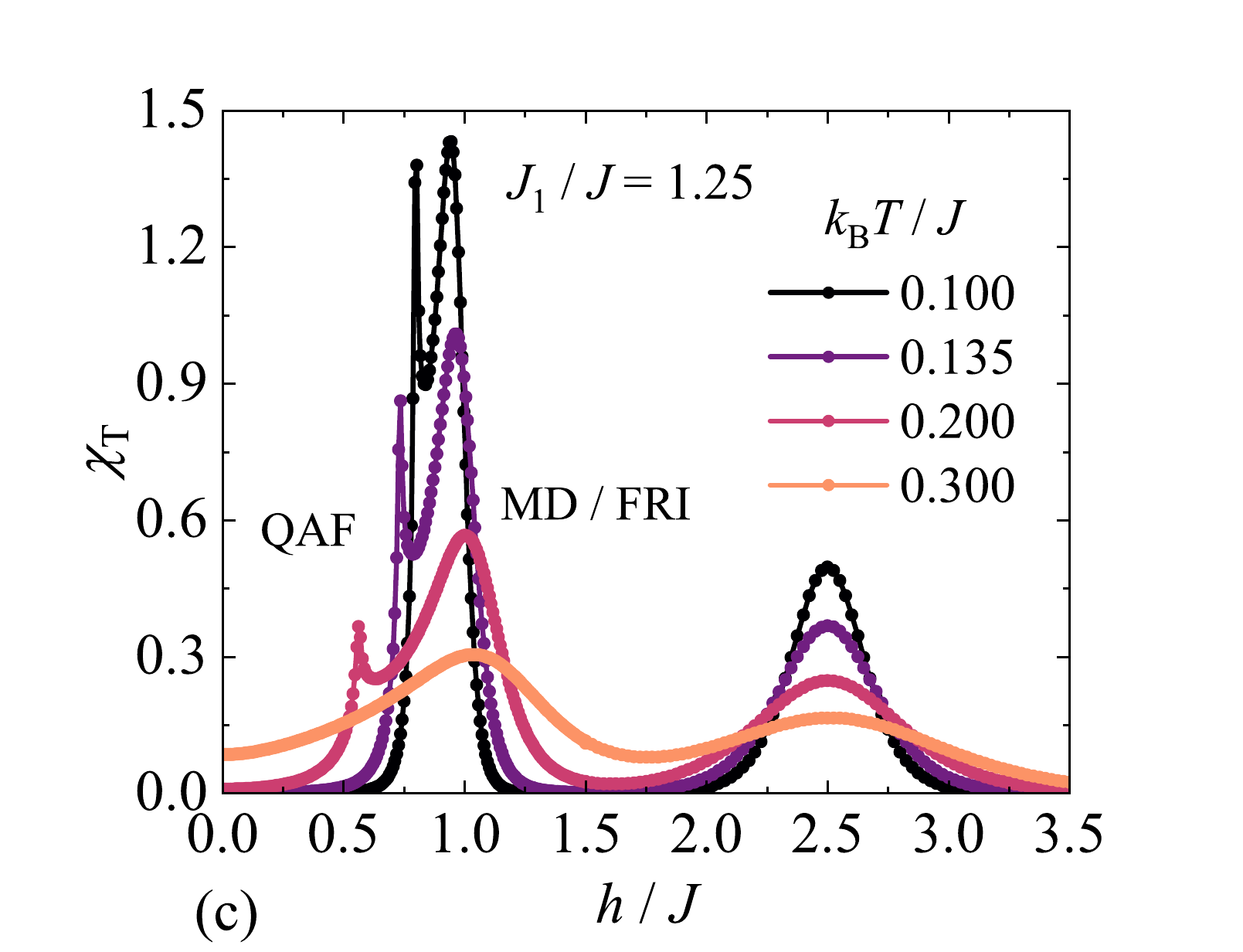}
\vspace*{-0.3cm}
\caption{Magnetic-field dependencies of the total magnetization (a), local magnetizations (b), and total magnetic susceptibility (c) of the spin-$1/2$ Ising-Heisenberg model on the extended Lieb lattice for the interaction ratio $J_1/J = 1.25$ at four different temperatures. Blue circles highlight the Ising-type critical points between the QAF and MD/FRI phases, which are for clarity omitted in the main panel for the local magnetization and displayed only in the inset.}
\label{fig7}
\end{figure*}

Next, let us provide an independent verification of the magnetic behavior of the spin-$1/2$ Ising-Heisenberg model on the extended Lieb lattice for the interaction ratio $J_1/J = 1.25$. At this fixed value of the interaction ratio, the corresponding finite-temperature phase diagram shown in Fig. \ref{fig5}(e) predicts the absence of thermal transitions between the MD and FRI phases and the existence of two distinct continuous QAF–MD and QAF–FRI phase transitions. Fig. \ref{fig7}(a) presents the isothermal magnetization curves at four different temperatures with blue circles denoting the Ising critical points, which correspond to the continuous thermal phase transitions. At the lowest considered temperature $k_{\rm B}T/J = 0.1$, the magnetization curve exhibits a continuous thermal phase transition between the QAF and FRI phases corresponding to two distinct intermediate plateaus at 0 and $3/5$ of the saturation magnetization, respectively.  Apart from this special critical point, the magnetization curve does not involve any additional singularity. Within the range of moderate temperatures $0.135 \lesssim k_{\rm B}T/J \lesssim 0.27$, one should instead expect the emergence of an intermediate $1/5$-plateau associated with the MD phase. This expectation follows from the sign change of the effective field occurring approximately at $k_{\rm B}T/J \approx 0.135$, which should favor the QAF–MD continuous thermal transition rather than the QAF–FRI transition. However, the corresponding magnetization curve from this temperature range (e.g. $k_{\rm B}T/J = 0.2$) does not clearly uncover the intermediate $1/5$-plateau indicating that a more detailed analysis is required. At sufficiently high temperatures (e.g., $k_{\rm B}T/J = 0.3$), the magnetization curve already becomes completely smooth and continuous throughout the entire magnetic-field range. Although no additional thermal phase transitions are observed, a crossover phenomenon can still be identified near $h/J \approx 2.5$ for all considered temperatures reflecting a residual signature of the zero-temperature discontinuous transition between the FRI and FM phases.

To provide more compelling evidence for the continuous thermal phase transitions between the QAF and MD phases, Fig. \ref{fig7}(b) shows the corresponding local magnetizations of the Ising (solid lines) and Heisenberg (dashed lines) spins. For clarity, blue circles indicating the critical points are omitted from the main panel and are only displayed in the inset. At sufficiently low temperatures (e.g., $k_{\rm B}T/J = 0.1$), the local magnetization of the Heisenberg spins initially remains nearly zero and then begins to increase gradually near the critical field after passing of which it starts to increase more rapidly and eventually reaches the saturation. Similarly, the local magnetization of the Ising spins also remains nearly zero at low magnetic fields, but then it begins to decrease gradually towards negative values in the vicinity of the critical field before it finally increases and ultimately reaches saturation at higher fields around $h/J \approx 2$ [see Fig. \ref{fig7}(b)]. As a result, the intermediate 0- and $3/5$-plateaus appear in the magnetization curves shown in Fig. \ref{fig7}(a). Increasing the temperature generally smears out the sharp changes in both local magnetizations and the critical point vanishes at sufficiently high temperatures (e.g. $k_{\rm B}T/J = 0.3$). A closer inspection of the local magnetization of the Ising spins provided in the inset of Fig. \ref{fig7}(b) reveals that at intermediate temperature $k_{\rm B}T/J = 0.135$ this local magnetization initially increases in a low-field regime reaching a small round maximum. This feature is a characteristic signature of the MD phase, which is accompanied with a positive effective field and, consequently, a positive value of the local magnetization of the Ising spins. The local maximum becomes even more pronounced with increasing temperature, as it is evident by comparing the curves for $k_{\rm B}T/J = 0.135$ and $0.2$, before it gradually diminishes upon further increasing the temperature (e.g., $k_{\rm B}T/J = 0.3$). Recall from the previous discussion for the particular case $J_1/J = 0.9$ that the local magnetization of the Ising spins indeed increases steadily in the vicinity of the continuous QAF-MD phase transition and then suddenly decreases near the discontinuous MD–FRI phase transition [see, for instance, the magnetization data for $k_{\rm B}T/J = 0.05$ in Fig. \ref{fig6}(b)].

The continuous thermal phase transitions between the QAF and MD/FRI phases are most clearly manifested in Fig. \ref{fig7}(c), which shows the isothermal field dependencies of the magnetic susceptibility. At the lowest temperature $k_{\rm B}T/J = 0.1$ considered in Fig. \ref{fig7}(c), the magnetic susceptibility displays a sharp peak near the critical field $h/J \approx 0.8$ signaling a continuous thermal phase transition, which is subsequently followed by two additional round maxima located around $h/J \approx 0.9$ and $2.5$. While the former sharp peak associated with the thermal phase transition  increases with system size and tends to diverge in the thermodynamic limit, the latter two round maxima remain finite. With increasing temperature, the sharp susceptibility peak ascribed to the thermal phase transition shifts towards lower magnetic fields, while its magnitude progressively diminishes until it completely disappears at sufficiently high temperatures (see for instance $k_{\rm B}T/J = 0.3$). The intermediate round maximum apparently reflects a crossover phenomenon between the MD and FRI phases. In addition, the last round maximum near $h/J \approx 2.5$ relates to a crossover phenomenon associated with the zero-temperature phase transition between the FRI and FM phases. 

\begin{figure}[t]
\centering
\includegraphics[width=0.35\textwidth]{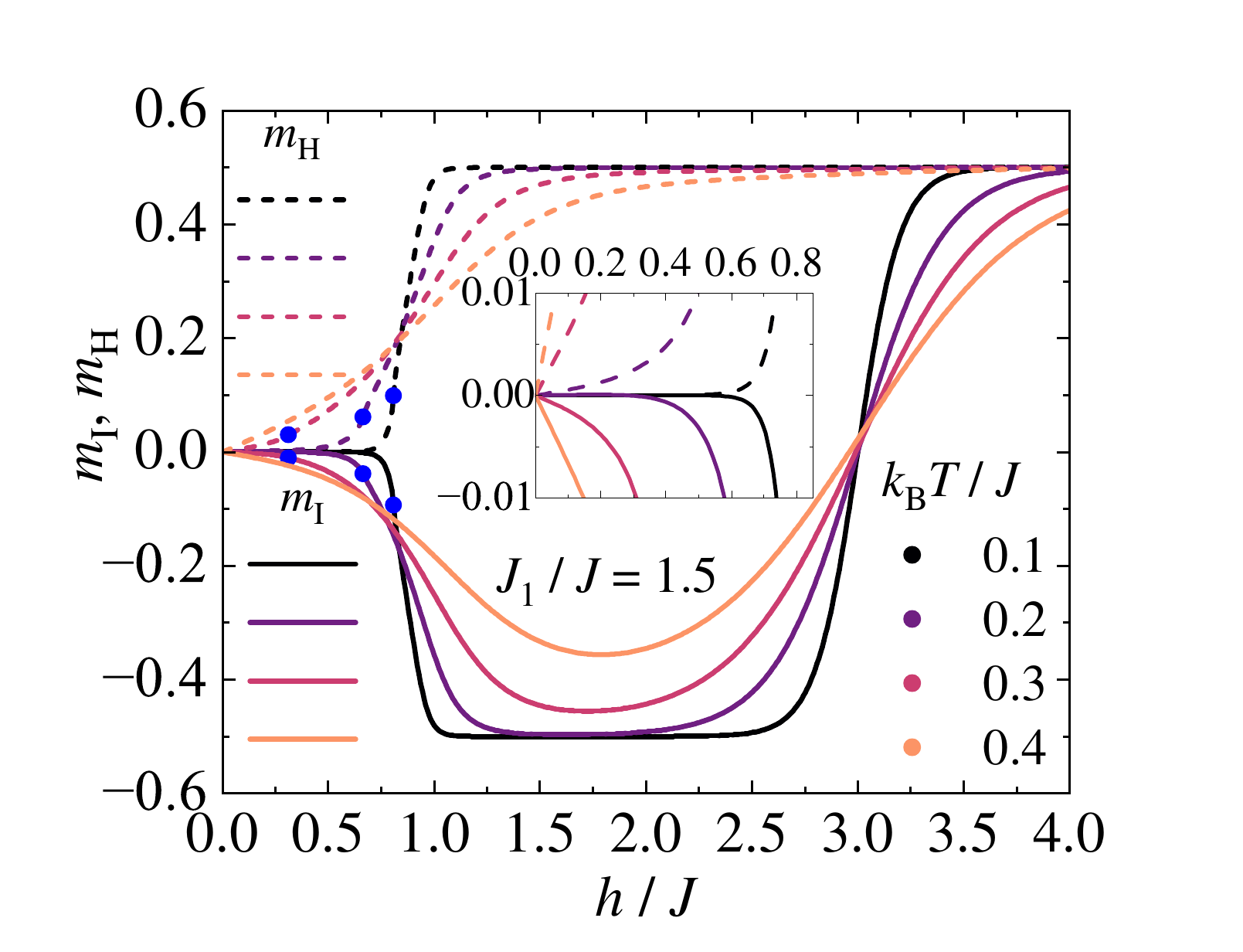}
\vspace*{-0.3cm}
\caption{Local magnetization of the Ising and Heisenberg spins in the spin-$1/2$ Ising-Heisenberg model on the extended Lieb lattice as a function of the magnetic field for the interaction ratio $J_1/J = 1.5$ at four different temperatures. Blue circles indicate positions of the Ising-type critical points.}
\label{fig8}
\end{figure}

Last but not least, let us focus our attention on the particular case with even higher value of the interaction ratio, for which the MD phase is progressively suppressed by the dominant FRI phase and eventually completely disappears from the global phase diagram. To illustrate this behavior, Fig. \ref{fig8} shows the magnetic-field dependencies of the local magnetization of the Ising and Heisenberg spins of the spin-$1/2$ Ising-Heisenberg model on the extended Lieb lattice for a fixed value of the interaction ratio $J_1/J = 1.5$ at four different temperatures with blue circles indicating positions of the Ising-type critical points. The magnetic behavior displayed in Fig. \ref{fig8} is fully consistent with the corresponding finite-temperature phase diagram shown in Fig. \ref{fig5}(f). The inset of Fig. \ref{fig8} further corroborates that the local magnetization of the Ising spins no longer exhibits any characteristic signature of the MD phase for such a large value of the interaction ratio. As a matter of fact, the local magnetization of the Ising spins shows no evidence of the initial field-induced rise associated with the MD phase at any temperature, but instead it exhibits immediate downturn from zero towards negative values characteristic of the FRI phase without developing any positive local maximum. This observation is fully consistent with the fact that the effective field of the corresponding spin-1/2 Ising square lattice remains negative in the respective range of magnetic fields and temperatures when the interaction ratio $J_1/J = 1.5$ is considered.

\section{Conclusion}
\label{conclusion}

The exact mapping of the spin-$1/2$ Ising-Heisenberg model on the extended Lieb lattice in a magnetic field onto an effective spin-$1/2$ Ising model on the square lattice was accomplished by means of the generalized decoration-iteration transformation. The magnetic properties of the original model were subsequently analyzed after solving the corresponding Ising model with an effective nearest-neighbor interaction and an effective field. Several exact analytical results were obtained for the parameter region characterized by vanishing effective field, which enabled a rigorous investigation of diverse discontinuous and continuous thermal phase transitions both in the presence and absence of a genuine magnetic field. Since an exact solution of the effective Ising model on the square lattice is available only for zero effective field, classical Monte Carlo simulations were additionally employed to examine thermal and magnetic properties in a more general case with particular emphasis on an independent verification of the predicted discontinuous and continuous thermal phase transitions. 

The ground-state phase diagram of the spin-$1/2$ Ising-Heisenberg model on the extended Lieb lattice comprises four distinct phases, namely, QAF, MD, FRI, and FM. The phase transitions between the MD and FRI phase remain confined to a parameter region characterized by a ferromagnetic effective interaction and a vanishing effective field both at zero as well as nonzero temperatures. Consequently, the dome-shaped surface of discontinuous MD-FRI thermal phase transitions as well as the line of the Ising critical points corresponding to the continuous MD-FRI thermal phase transitions bounding this surface from above were determined exactly. On the other hand, the continuous QAF–MD and QAF–FRI thermal phase transitions were determined approximately and subsequently independently verified by Monte Carlo simulations, which unambiguously confirmed the existence of both continuous thermal phase transitions. All numerical results extracted from the Monte Carlo simulations are in excellent agreement with the exact as well as approximate analytical predictions. This finding demonstrates that the approximation used for determining the global phase diagram is sufficiently accurate. 

It may be concluded that the exact and approximate analytical results obtained for the spin-$1/2$ Ising-Heisenberg model on the extended Lieb lattice in a magnetic field provide valuable insight into the underlying physics of both continuous and discontinuous thermal phase transitions. This observation is particularly important because one may anticipate similarly rich, albeit more complex, behavior in the fully quantum Heisenberg model on the extended Lieb lattice. Such an expectation is supported by the recent observation of discontinuous and continuous thermal phase transitions in the spin-$1/2$ Heisenberg model on the diamond-decorated square lattice in a magnetic field \cite{cac23}, whose nature and origin are fully consistent with the corresponding thermal phase transitions reported for the analogous spin-$1/2$ Ising-Heisenberg model on the diamond-decorated square lattice in a magnetic field \cite{str23}. Whether a similar correspondence also exists for the extended Lieb lattice will be investigated by the present authors in a separate study.

\section*{Acknowledgements}
The authors acknowledge funding by the grant of Slovak Research and Development Agency under the contract No. APVV-24-0091 and by the grant of The Ministry of Education, Research, Development and Youth of the Slovak Republic under the contract No. VEGA 1/0695/23.

\end{document}